\documentclass{nature}

\usepackage{graphicx}
\usepackage{amsmath}	
\usepackage{amssymb}
\usepackage{bm}	
\usepackage{lipsum}
\usepackage{afterpage}
\usepackage{multirow}
\usepackage{makecell}
\usepackage{xcolor}
\usepackage{caption}
\usepackage{longtable}
\usepackage{booktabs}

\usepackage{float}

\usepackage[normalem]{ulem}

\usepackage{lineno}

\newcommand{\src}{4U~1630$-$472}

\title{Evidence of mutually exclusive outflow forms from a black hole X-ray binary}

\author{Zuobin Zhang$^{1,2}$, Jiachen Jiang$^{3}$, Francesco Carotenuto$^{2,4}$, Honghui Liu$^{5,1}$, Cosimo Bambi$^{1,6,\dag}$, Rob P. Fender$^{2,7}$, Andrew J. Young$^8$, Jakob van den Eijnden$^{9}$, Christopher S. Reynolds$^{10}$, Andrew C. Fabian$^{11}$, Julien N. Girard$^{12}$, Joey Neilsen$^{13}$, James F. Steiner$^{14}$, John A. Tomsick$^{15}$, St\'ephane Corbel$^{16}$ and Andrew K. Hughes$^{2}$}

\begin{document}

\maketitle

\begin{affiliations}
 \item Center for Astronomy and Astrophysics, Center for Field Theory and Particle Physics, and Department of Physics, Fudan University, Shanghai 200438, China
 \item Astrophysics, Department of Physics, University of Oxford, Keble Road, Oxford OX1 3RH, UK
 \item Department of Physics, University of Warwick, Gibbet Hill Road, Coventry CV4 7AL, UK
 \item INAF-Osservatorio Astronomico di Roma, Via Frascati 33, I-00078, Monte Porzio Catone (RM), Italy
 \item Institut f\"ur Astronomie und Astrophysik, Eberhard-Karls Universit\"at T\"ubingen, D-72076 T\"ubingen, Germany
 \item School of Natural Sciences and Humanities, New Uzbekistan University, Tashkent 100007, Uzbekistan
 \item Department of Astronomy, University of Cape Town, Private Bag X3, Rondebosch 7701, South Africa
 \item School of Physics, Tyndall Avenue, University of Bristol, Bristol BS8 1TH, UK
 \item Anton Pannekoek Institute for Astronomy, Universiteit van Amsterdam, Science Park 904, 1098, XH, Amsterdam, The Netherlands
 \item Department of Astronomy, University of Maryland, College Park, MD 20742-2421, United States
 \item Institute of Astronomy, University of Cambridge, Madingley Road, Cambridge CB3 0HA, UK
 \item LIRA, Observatoire de Paris, Universit\'e PSL, Universit\'e Paris Cit\'e, Sorbonne Universit\'e, CY Cergy Paris Universit\'e, CNRS, 92190 Meudon, France
 \item Villanova University, Department of Physics, Villanova, PA 19085, United States
 \item Center for Astrophysics, Harvard \& Smithsonian, Cambridge, MA 02138, United States
 \item Space Sciences Laboratory, 7 Gauss Way, University of California, Berkeley, CA 94720-7450, United States
 \item Universit\'e Paris Cit\'e and Universit\'e Paris Saclay, CEA, CNRS, AIM, F-91190 Gif-sur-Yvette, France
 \item[] $^\dag$ Corresponding author: bambi@fudan.edu.cn
\end{affiliations}

\begin{abstract}

Accretion onto black holes often leads to the launch of outflows that significantly influence their surrounding environments. The two primary forms of these outflows are X-ray disk winds—hot, ionized gases ejected from the accretion disk—and relativistic jets, which are collimated streams of particles often expelled along the rotational axis of the black hole. While previous studies have revealed a general association between spectral states and different types of outflows, the physical mechanisms governing wind and jet formation remain debated. Here, using coordinated NICER and MeerKAT observations of the recurrent black hole X-ray binary \src, we identify a clear anti-correlation between X-ray disk winds and jets: during three recent outbursts, only one type of outflow is detected at a time. Notably, this apparent exclusivity occurs even as the overall accretion luminosity remains within the range expected for a standard thin disk, characteristic of the canonical soft state. These results suggest a competition between outflow channels that may depend on how the accretion energy is partitioned between the disk and the corona. Our findings provide new observational constraints on jet and wind formation in X-ray binaries and offer a fresh perspective on the interplay between different modes of accretion-driven feedback.


\end{abstract}

Black hole X-ray binaries (BH~XRBs) consist of a stellar-mass black hole that accretes material from a donor star. X-ray observations of BH~XRBs have identified two main accretion states, the spectrally hard state and the soft state\cite{Remillard2006}. In the hard state, the X-ray spectrum is dominated by thermal Comptonization emission from a hot corona near the black hole\cite{Thorne1975, Sunyaev1979}. In the soft state, coronal emission is suppressed and the X-ray spectrum is dominated by thermal emission originating from the accretion disk\cite{Novikov1973, Shakura1973}. The difference between the hard and soft states is also manifested in the X-ray temporal behavior of the source, including the emergence and disappearance of quasi-periodic oscillations and changes in the level of aperiodic variability\cite{Homan2005, Belloni2010}.

Outflows in BH XRBs typically show two forms: disk winds and jets\cite{Gallo2005, Russell2013, Bright2020}. Disk winds often exhibit multiple phases, with less ionized gas observed in the optical and UV bands and more ionized gas detected in the X-ray band\cite{Munoz2019, Munoz2016, Munoz2022, Neilsen2023, Zhang2024}. Detected as blue-shifted absorption lines by highly ionized plasma in the X-ray spectra, X-ray winds are more prevalently detected in the soft state of high-inclination systems and are often envisioned as biconical, roughly equatorial outflows with velocities ranging from a few hundred to a few thousand kilometers per second\cite{Ponti2012, Miller2008}. Recently, observations at longer wavelengths have revealed the existence of lowly ionized disk winds\cite{Munoz2016, Munoz2019, Sanchez2020, Castro2022}. For \src, the optical counterpart has yet to be identified, and consequently, the presence and properties of any low-ionization disk winds remain unknown. In this study, we therefore focus exclusively on the X-ray disk winds.

Jets represent powerful, collimated, and relativistic outflows that produce electromagnetic radiation always peaking in the radio or infra-red band via synchrotron processes\cite{Fender2004, Fender2006, Fender2014}. Previous simultaneous X-ray and radio observations have illustrated that the accretion state of a BH~XRB system determines the characteristics of its outflows. Collimated, partially self-absorbed compact jets are frequently found in the hard state\cite{Fender2004, Fender2006}. In contrast, in the soft state, compact jets appear strongly suppressed, while powerful X-ray disk winds appear\cite{Russell2011, Miller2012, Neilsen2009, Fender1999_2}.

If X-ray disk winds and jets appear and disappear along with state transitions, it is natural to wonder whether there is a connection between X-ray winds and jets. Two possible models have been proposed: one explanation attributes the observed anti-correlation between X-ray disk winds and jets to variations in wind density, which may influence the magnetic field configuration in the inner disk \cite{Miller2012}; another interpretation considers the state-dependent nature of the form of outflows as evidence of a material tradeoff, where mass loss through winds from the outer disk reduces the matter available to fuel jets from the inner disk \cite{Neilsen2009}. However, it is difficult to test each scenario as observations show complex variability patterns in X-ray disk winds, jets, and disk accretion rates\cite{Lee2002, Miller2006}.

Motivated by the search for a connection between jets and winds, we investigated the variability of the X-ray disk winds and jets in the soft state of \src, a recurrent X-ray transient considered to be a black hole candidate based on its X-ray spectral and timing behavior\cite{Jones1976, Priedhorsky1986, Kuulkers1997, Remillard2006}. We focused on its latest three outbursts, which were intensively monitored by the Meer Karoo Array Telescope (MeerKAT) and the Neutron Star Interior Composition Explorer (NICER) from 2020 to 2023. We measured the properties of the X-ray winds using spectral data from NICER\cite{Neilsen2018}. The radio luminosity was measured by MeerKAT, which serves as an indicator of jet activity. Figure~1 shows two of the twenty-eight simultaneous NICER and MeerKAT observations as examples. One of the NICER observations shows significant evidence of a blue-shifted Fe \textsc{xxvi} Ly$\alpha$ line ($\sim 7.0$~keV; $> 5 \sigma$), suggesting X-ray disk winds with a line-of-sight velocity of $\sim 300$~km/s directed towards the observer. The unabsorbed X-ray flux during this observation is $1.6 \times 10^{-8}$~erg~cm$^{-2}$~s$^{-1}$ in the 1–10~keV band. The simultaneous MeerKAT observation detects no significant radio emission with an upper limit of $<0.45$~mJy. In contrast, the other set of observations, also in a comparable soft state with a similar unabsorbed X-ray flux, shows no evidence of winds but strong evidence of synchrotron radio emission from jets with a flux density of $1.90 \pm 0.04$~mJy.

The NICER observations suggest that \src\ spent most of its time in a soft state during our observing periods: the X-ray spectrum is dominated by the thermal emission of a geometrically thin accretion disk with an Eddington ratio of $4\%-33\%$ (1--10~keV; assuming a black hole mass of 10~$M_{\bigodot}$ and a distance of 10~kpc, see Methods for details); the power-density spectrum shows little variability. We use simple Gaussian models to estimate the detection significance of the wind's absorption lines and then a photoionized plasma model to estimate the wind parameters, e.g., column density, ionization, and velocity (see Methods for more details). To study the jets of \src, we extract the 1.28~GHz radio flux density from the MeerKAT observations and conduct spectral analysis for the observations with a significant detection above 1~mJy. 

Our observations show clear evidence of a negative correlation between the column density of X-ray winds and radio flux density, indicating mutually exclusive forms of outflows in \src. Figure~2 illustrates the temporal evolution of the X-ray wind’s column density and the radio flux density, the shaded region indicating the period when the source was in the soft state. Generally, strong radio emission was detected during the soft state in 2021, with a flux density between 2 and 5~mJy, while the the X-ray spectra exhibited little or no evidence of X-ray winds ($N_{\rm H}<10^{22}$\,cm$^{-2}$). In contrast, the soft state in 2022-2023 shows weak radio emission ($<1$~mJy), but accompanied by significant evidence of X-ray winds, with column densities ranging from $3 \times 10^{22}$\,cm$^{-2}$ to $30 \times 10^{22}$\,cm$^{-2}$. No X-ray absorption lines or significant radio emission were detected during specific periods. 

Note that for the 2021 outburst period of interest, we find that the radio flux density is positively correlated with the strength of the corona component, supporting the scenario in which the underlying jet activity is directly driven by the accretion process. In this period, the corona flux varies between $3 \times 10^{37}$~erg~s$^{-1}$ and $1 \times 10^{38}$~erg~s$^{-1}$ (1--10~keV), and its proportion of total flux varies roughly between 17\% and 45\%. The source was in a disk-dominated state during this period (see Methods for further details).

Disk winds are often thought to dominate the mass-loss budget in XRBs\cite{Ponti2012, Neilsen2011}, but this is not yet firmly established. Recent work on MAXI~J1820+070, for example, suggests that jets may carry significant mass densities to propagate to the observed distances\cite{Savard2025}. Based on the observed radio luminosity of \src\ and minimum-energy condition, we estimate a jet mass-loss rate of $\sim 10^{18}$~g~s$^{-1}$ (see Methods), comparable to disk wind estimates. This suggests that both components may contribute non-negligibly to the outflow mass-loss, and the tradeoff between X-ray disk winds and jets is regulated by mass conservation. The results of GRS~1915+105 also suggest thermally-driven winds could deplete the mass of the disk enough to suppress relativistic jets\cite{Neilsen2009}.

All of our soft-state observations suggest that the accretion disk of \src\ remains consistent with a standard geometrically thin accretion disk\cite{Novikov1973}. The inner radius of the accretion disk is compatible with the innermost stable circular orbit. The significant evidence of radio jet emission in some of our soft-state observations suggests that such jets in \src\ are unlikely related to the build-up of strong magnetic fields in a geometrically thin, low accretion rate disk truncated at a large radius as in other low mass BH XRBs (e.g., MAXI~J1820+070\cite{You2023}). Instead, the jets and X-ray winds in \src\ may be launched from disk through the coupling between a spiral density wave in the disk and a Rossby vortex \cite{Tagger1999}. An alternative interpretation is that the configuration and strength of magnetic fields take charge of the exclusiveness between X-ray winds and jets. Strong poloidal magnetic fields can suppress the launch of magneto-thermal X-ray winds\cite{Waters2018}, while they are always required to accelerate and collimate jets\cite{Meier2001}. Thus, different configurations and strengths of magnetic fields may lead to different forms of outflows.

Previous observations of low-mass XRBs have revealed complex jet and wind behaviors, particularly in systems undergoing frequent state transitions on short timescales, such as neutron star Z sources and GRS~1915+105. Simultaneous X-ray and radio monitoring has shown that ionized disk winds and radio jets can coexist in sources like GX~13+1\cite{Rogantini2025, Homan2016}. In GRS~1915+105, long-term observations indicate that jet ejection events are associated with state transitions between discrete soft-state regimes\cite{Inoue2022, Klein2002}. In contrast, \src\ remains persistently below 30\% of the Eddington limit throughout the analyzed periods, consistent with a standard thin disk and showing no evidence for soft-state transitions. The jets observed in \src\ are therefore distinct from those associated with state changes in sources like GRS~1915+105 (see Methods for further details). Furthermore, direct comparisons with GX~13+1 are not straightforward, as GX~13+1 accretes near or above the Eddington limit, where radiation pressure likely plays a key role in the launch of disk winds\cite{Rogantini2025}.

This is clear evidence of mutually exclusive forms of outflows in a BH~XRB. While the accretion disk remains constant in accretion rate, the form of outflows in this object switches between jets, traced by strong radio emission, and X-ray disk winds, traced by X-ray absorption lines. These results offer fundamental new insights into the long-term coupling of accretion disk and outflows around black holes and suggest compelling evidence for self-regulation mechanisms in the growth of stellar-mass black holes.


\begin{figure}
       \centering
       \vspace{-0.5cm}
       \includegraphics[width=0.9\linewidth]{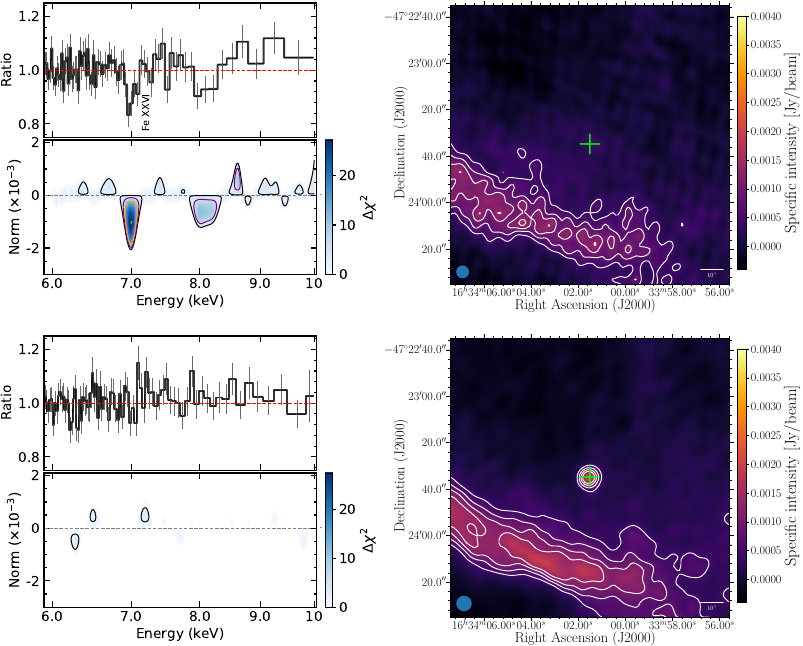} 
       \\
       \vspace{-0.5cm}
       \label{wind_jet_example}
\end{figure}

\newpage

\noindent
\textbf{Figure~1\boldmath{$|$} Two examples of simultaneous NICER and MeerKAT observations of \src}. We use a $\sim$1~ks segment of the top observation to ensure that two NICER observations (ObsIDs: 6130010111 and 4130010124) have same duration, and the signals are comparable. The first NICER observation shows evidence of disk wind (absorption line at $\sim 7.0$~keV ($> 5 \sigma$); top left) and the other without disk wind (bottom left); two (quasi-)~simultaneous radio images are shown at the right, one without a radio jet (top right) and one with a radio jet (bottom right). The top panels of the X-ray spectral figures show the residuals to the absorbed disk+corona model, with error bars representing 1$\sigma$ confidence intervals on the data-to-model ratio; and the bottom panels show the results of a line search over 6-10~keV band. In the radio images, the contours are at 3, 5, 7, 9 times the noise RMS, which is 150 $\rm{\mu}Jy$ for both radio images. The cross represents the source position, while the feature in the bottom-left corner is the emission of the Galactic plane. The blue circle on the bottom left is the MeerKAT synthesized beam. Colour bars indicate the specific intensity ($\rm Jy$~${\rm beam}^{-1}$) for the radio maps and $\Delta \chi^2$ values for blind line search.


\newpage


\begin{figure}
       \centering
       \includegraphics[width=1.02\linewidth]{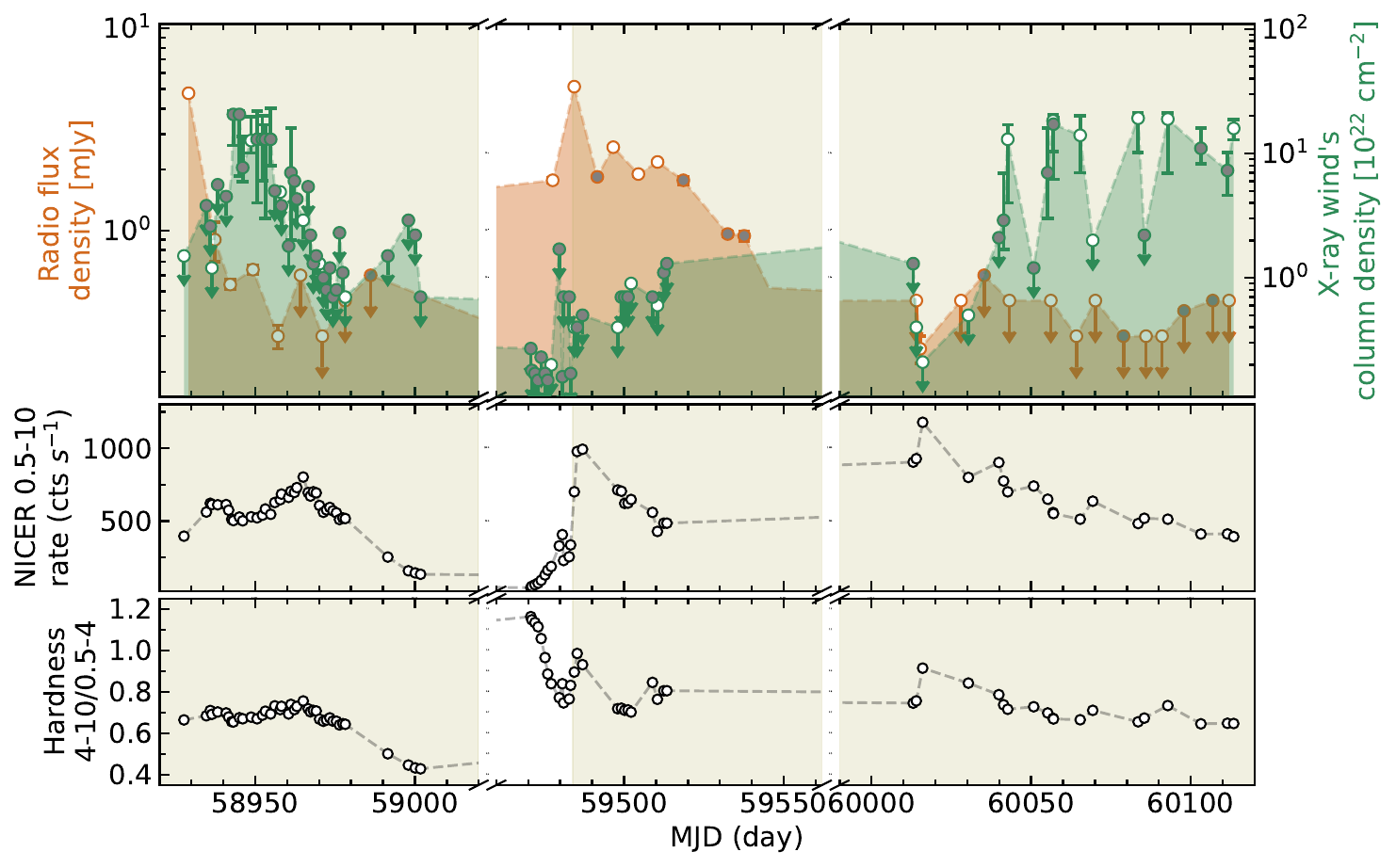}  
       \\
       \vspace{-0.3cm}
       \label{evolution_wind_radio}
\end{figure}

\newpage

\noindent
\textbf{Figure~2\boldmath{$|$} The evolution of the X-ray winds and radio jets for the 2020, 2021 and 2022-2023 outbursts.} Top panel: evolution of the X-ray wind’s column density and the flux density of radio emission over time, with the tan-shaded area delegating the period when the source is in the soft state. Dark-yellow points show the radio flux density with 1$\sigma$ confidence interval error bars, and 3$\sigma$ upper limits for non-detections. Green points show the X-ray wind column density with 3$\sigma$ confidence intervals for detections and 3$\sigma$ upper limits for non-detections. The open points indicate observations where X-ray and radio emissions occurred simultaneously or quasi-simultaneously ($\Delta$MJD$<$2), while the filled points represent data from other observations. Central panel: the $0.5-10$~keV NICER light curve. Bottom panel: the corresponding NICER hardness, which is defined as the ratio between count rates in $4-10$~keV and $0.5-4$~keV bands. Error bars, which are too small to be visible for most data, represent 1$\sigma$ confidence interval.


\newpage

\begin{methods}

\subsection{NICER data reduction.}

The NICER observations we study in this work are listed in Supplementary Table~1. They are processed using the NICER Data Analysis Software (NICERDAS; version 2021-04-01\_V008) and the NICER Calibration Database (CALDB; version 20210707). Standard data reduction is carried out with the task nicerl2, adopting the default filtering criteria: a pointing offset smaller than $54^{\prime \prime}$, an angular distance of more than $40^{\circ}$ from the bright Earth limb and $30^{\circ}$ from the dark Earth limb, and exclusion of time intervals when the satellite was within the South Atlantic Anomaly. Events flagged as “undershoot” or “overshoot” (EVENT\_FLAGS = bxxxx00) and forced triggers (EVENT\_FLAGS=bx1x000) are removed. In addition, data from detectors \#14 and \#34 are excluded due to electronic noise. Background spectra are generated using the nibackgen3C50 tool\cite{Remillard2022}. We use the nicerl3-spect task to produce the source spectrum together with the corresponding redistribution matrix file (RMF) and ancillary response file (ARF) for each observation. All spectra were grouped to ensure a minimum of 20 counts per energy bin.

\subsection{MeerKAT data reduction.}

We observed \src\ with MeerKAT as a part of the ThunderKAT large survey project\cite{Fender2018}. The source was observed with a weekly cadence for 39 epochs between 2020 and 2023. A summary of the MeerKAT observations is given in Supplementary Table~2. Each observation consisted of a single 15-minute on-source scan preceded and followed by a 2-minute scan of a nearby gain calibrator (J1726-5529). PKS~1934-638 (J1939-6342) was used for flux and bandpass calibration. Our observations were conducted at L-band, with a central frequency of 1.284~GHz and a total bandwidth of 856~MHz. Data are processed using oxkat\cite{Heywood2020}, a set of scripts developed for automatically processing MeerKAT continuum data. Within oxkat, the calibration is performed using CASA\cite{CASATeam2022}, radio frequency interference (RFI) is removed using both CASA and TRICOLOUR\cite{Hugo2022}, and the direction-independent self-calibration relies on CUBICAL\cite{Kenyon2018}. Moreover, imaging and deconvolution are performed with WSCLEAN\cite{Offringa2014} using a Briggs weighting scheme with a robust parameter of -0.3. Due to the presence of extended diffuse emission, to which MeerKAT is particularly sensitive, we create an ad-hoc thresholded mask with BREIZORRO\cite{Ramaila2023} at 6.5 times the image RMS noise and we use it to perform a deeper deconvolution after the first imaging step. This process allows us to obtain a rms-noise level ranging between 100 and 200~$\mu$~Jy. We fit a point source in the image plane with the CASA task IMFIT to obtain the radio flux density for each epoch. In the case of non-detection, upper limits are reported at three times the rms noise level in a source-free region covering the target position. For the radio spectral index $\alpha$, we divide each epoch in 8 spectral chunks with a 107 MHz bandwidth between 856~MHz and 1.71~GHz, imaging each of them and obtaining the target flux density $S_{\nu}$ with the process mentioned above. $\alpha$ is then obtained by fitting the data for a power-law $S_{\nu} \propto \nu^{\alpha}$. For all epochs, we exclude the first spectral range between 856 and 963 MHz due to a higher level of data flagging ($\sim$70\%, due to RFI) with respect to the average in the MeerKAT band (roughly $\sim$30\%). The extracted radio flux density and the radio spectral index for the entire monitoring are reported in Supplementary Table~2. 

\subsection{Significance test of the absorption features in NICER spectra.}

We start fitting NICER continuum spectra with the following phenomenological model: tbabs $\times$ xscat $\times$ edge $\times$ (cflux $\times$ diskbb + cflux $\times$ nthcomp). Tbabs is included to account for absorption by Galactic interstellar medium. Diskbb accounts for the thermal emission from the optically thick accretion disk. In some cases, a weak nthcomp is needed to model the power-law component from the corona. The model xscat accounts for the scattering of photons from interstellar gas and dust grains. A dust-scattering halo is common for sources with a high hydrogen column density ($N_{\rm H}$) along the line of sight. Kalemci et al. studied the effects of dust scattering in the case of \src\ \cite{Kalemci2018}. The xscat model has the following parameters, $X_{\rm pos}$, $N_{\rm H}$ and $R_{ext}$, which indicate the position of the dust scattering halo, the hydrogen column density of the scattering source and the radius of the circular extraction region, respectively. In our fitting, we fix $X_{\rm pos} = 0.9$\cite{Kalemci2018, Connors2021} and $R_{ext}  = 180^\prime$. $N_{\rm H}$ is tied to the line-of-sight hydrogen column given by the tbabs model. The residuals at $\sim 2.4$~ keV are attributed to the known effective area calibration uncertainty of the instrument, particularly near the gold M edge due to the reflector's gold coating\cite{Wang2020}. We use an absorption edge model to model the feature around $\sim 2.4$~keV empirically. Note that in some observations,  the spectra show evidence of a broad peak around 6.4~keV, which indicates the existence of a reflection component by the optically thick accretion disk surrounding the black hole\cite{Ross2005, Ross2007, Garcia2014}. We introduce a Gaussian model to account for the reprocessed Fe-K emission from the disk with the centroid energy E$_{\rm line}$ fixed to 6.4~keV. The cflux model is included for each component to calculate the unabsorbed flux of each component in the $1.0-10.0$~keV band.

Supplementary Figure~1, as an example, shows the unfolded NICER spectrum and the fitting residuals of the two observations mentioned in Figure~1. Based on the ratio of the spectra to this model fit, we note the presence of an absorption line feature in some spectra at $\sim 7$~keV, consistent with Fe XXVI absorption from the known X-ray disk winds. Following on from this, we perform a blind line search over $6-10$~keV, stepping a Gaussian line (varying $\sigma$ and the normalization, allowed to be positive or negative) and recording the $\Delta \chi^2$ at each point of this grid (Figure~1, left). Comparing the change in $\chi^2$, we distinguish the observations in which the addition of the Gaussian absorption line is significant ($3 \sigma$ or greater). With this phenomenological model, we calculate the disk and corona component flux in the $1.0-10.0$~keV band, shown in panel (c) in Supplementary Figure~2.

\subsection{Eddington Ratio Estimation.}

\src\ is located toward the Galactic center, obscured by a large neutral hydrogen column\cite{Parmar1995, Gatuzz2019} ($N_{\rm H} \simeq 10^{23}$~cm$^{-2}$). This heavy absorption makes identifying its optical counterpart challenging, despite the latter's importance for constraining the primary accretor's mass and distance\cite{Kuulkers1997}. While the black hole's dynamical mass remains undetermined, an indirect estimate of $\sim 10$~M$_{\bigodot}$ was derived by correlating the photon index with the accretion rate\cite{Seifina2014}. Infrared observations during the 1998 outburst suggest a distance of 10–11 kpc\cite{Augusteijn2001, Seifina2014}, consistent with measurements from the source's dust-scattering halo\cite{Kalemci2018}.

We calculate the Eddington ratio of \src\ using the unabsorbed flux in the 1--10~keV band. Our results show that the source's total soft state luminosity ranges from $\sim 5 \times 10^{37}$ and $\sim 4 \times 10^{38}$~erg~s$^{-1}$, corresponding to Eddington ratios of 4\%--33\%. The highest luminosities were recorded in the first four NICER observations of the 2022–2023 outburst. During this period (MJD~$\approx$~59990–60030), the MAXI counts rate reached exceptionally high values, comparable to those observed at the peak of the 2012–2013 outburst \cite{Gatuzz2019}. This behavior represents a high flux phase that \src\ occasionally enters during mega outburst\cite{Abe2005}. Previous studies have suggested that the source in such epochs likely transitions into a steep power-law \cite{Hori2014} or an "anomalous" state \cite{Diaz2014} rather than the canonical soft state. In fact, during these four high-luminosity observations, neither radio jets nor disk winds were detected. As these data do not directly affect our main conclusions regarding winds–jets connections, we do not investigate them further. Excluding these non-standard-state observations, as well as those not relevant to the jets–winds comparison, the luminosity varies between 12\% and 26\% in terms of Eddington luminosity. When considering a broader energy band (0.01--100~keV), the Eddington fraction are between 18\%--33\%. We note that the black hole mass and distance remain uncertain, and thus these Eddington fractions should be regarded as approximate values. While the calculated luminosities correspond to sub-Eddington values under the assumed parameters, the possibility of the source entering a non-standard accretion state at comparable luminosities cannot be completely ruled out.

\subsection{Identification of spectral states.}

Supplementary Figure~2 shows the long-term X-ray monitoring of the source. Panel (a) and (b) depict Monitor of All-sky X-ray Image (MAXI) Gas Slit Camera (GSC) light curves in the $2--20$~keV bands and the corresponding hardness ($4--10$/$2--4$~keV), respectively.  The right panel (panel (d)) shows the MAXI hardness intensity diagram (HID) of \src. 

We classify the accretion state of \src\ using its spectral and temporal properties. The thermal disk emission contributes at least 70\% of the total luminosity in the NICER band (1--10 keV). The power density spectra (PDS) show no significant variability power - with root-mean-square normalization consistent with zero - and no quasi-periodic oscillations, consistent with typical soft-state characteristics\cite{wilkinson09,Remillard2006}.

Applying these criteria, we exclude NICER observations from September 13–26, 2021 when the source was in the hard state with Comptonization-dominated X-ray emission. The soft-state intervals (shaded region in Figure~2) show disk-dominated spectra with intrinsic luminosities of $\sim 4 \times 10^{37}$ and $\sim 3 \times 10^{38}$~erg~s$^{-1}$, accounting for 70–99.9\% of the total luminosity in the NICER band (1--10\,keV). Notably, during the 2021 radio outburst peak (maximum jet activity), the disk luminosity was $1.5 \times 10^{38}$ erg  s$^{-1}$, consistent with other epochs, although the coronal flux was slightly higher at this time. The PDS analysis (Supplementary Figure~6) shows no significant variability in any observation.

\subsection{Fitting with photoionized absorption model.}

For a more physical approach, we calculate a grid model with the XSTAR (version 2.2) code.  We assume that the illumination of the photoionized absorber is due to the absorption-corrected full continuum. For each observation, we extract the best fitting continuum model from the preceding phenomenological fit. We average the best-fit continuum model for each observation to get an averaged continuum model. It is used as the ionizing continuum for the photoionized absorber to calculate the XSTAR table model, which includes three free parameters: the absorber column density ($N_{\rm H}$), the ionization parameter ($\log \xi$) and the blue-shift because of outflow velocity ($v$). Thus the final model is: tbabs $\times$ xscat $\times$ edge $\times$ XSTAR $\times$ (cflux $\times$ diskbb + cflux $\times$ nthcomp). For the observations in which no significant absorption lines were detected in the previous fitting, i.e., with a significance $< 3\sigma$, we adopt the following strategy: firstly, we select the observation closest in time and with significant absorption lines detected as the reference observation; the blue-shift of the target observation is simply fixed at the corresponding value of the reference observation; for the ionization parameter, we get a predicted value with the formula $\xi = L_{\rm X} / (nr^2)$, assuming the $nr^2 = \rm constant$ in the transition between observations; the column density is treated as a free parameter, and typically an upper limit be obtained. The fitting results are reported in Supplementary Table~3. 

\subsection{The origin of the optically thin radio emission. }

Radio jets in BH XRBs typically manifest in two forms, both producing synchrotron radiation that peaks in either the radio or infrared bands \cite{Fender2004, Fender2006}. The optically thick emission, characterized by a spectral index $\alpha \gtrsim 0$ (following $S_{\nu} \propto \nu^{\alpha}$), is commonly observed during the hard state \cite{Corbel2000, Fender2001, Fender2004} and indicates the presence of steady compact jets. In contrast, optically thin emission ($\alpha < 0$) is typically associated with transient jets as discrete ejections \cite{Mirabel1994, Hjellming1995, Tingay1995}, which often show rapid variability \cite{Hjellming1995, Fender1999, Fender2001, Russell2013}.

During the 2021 soft state outburst, \src\ exhibited radio spectral indices between $-1.4$ and $-0.5$ (Supplementary Table~1), consistent with optically thin emission from transient jets with discrete ejections. Similar jet activity in disk-dominated states of \src\ in 2021 has been observed in other BH XRBs. In 4U~1543--47, significant radio flares were detected during the disk-dominated state\cite{Zhangx2025}, reaching 14~mJy at 1.28~GHz measured by MeerKAT. VLA observations of EXO~1846--031 revealed comparable soft-state radio flares (11~mJy at 5.5~GHz) with sub-arcsecond resolution\cite{Williams2022}; XTE~J1752--223 also exhibited jet activity in its disk-dominated state\cite{Brocksopp2013}. For \src, the typical 1.28~GHz flux density of 3~mJy and spectral index of $\alpha=-0.8$ yield a jet luminosity ($1-100$~GHz) of 0.8\% of the Eddington luminosity, consistent with other BH XRBs.

Notably, X-ray disk winds have not been detected in any radio-active soft-state observations of these sources\cite{Parra2024}—even in high-inclination systems like EXO~1846--031\cite{Draghis2020}, where winds might otherwise be expected. This supports our proposed mechanism of mutual exclusivity between jets and disk winds in the soft state.

\subsection{The positive hard X-ray and radio luminosity correlation.}

The observed radio emission is considered to originate from synchrotron radiation produced by jet material moving away from the central compact object. In the case of transient jet, a single event produces a flare with a characteristic duration ranging from about an hour to a day \cite{Mirabel1998, Brocksopp2002}. The evolution of such a flare can be described by the expanding synchrotron bubble model \cite{vanderLaan1966}, in which non-thermal electrons radiate as the ejected plasma sphere expands and moves outward. Based on this framework, we analyze the NICER X-ray data taken within a time window of $\pm$2 days around each radio observation to probe the central engine state during transient jet ejection.

The radio observations were made for $\sim$15 minutes at roughly weekly intervals, resulting in sparse temporal coverage. Nevertheless, during the first five radio observations of the 2020 outburst and all ten observations of the 2021 outburst, positive radio fluxes were detected consecutively, suggesting that the source remained in a radio-active phase for a high fraction of time. Such a high duty cycle may arise from either a series of short-lived transient flares produced by repeated jet ejections, as seen in GRS~1915+105\cite{Fender1999}, or prolonged radio emission caused by ejecta-ISM interactions \cite{Russell2019}.

A clear distinction between these two scenarios is found when comparing the 2020 and 2021 outbursts, as shown in Supplementary Figure~3. During the 2021 soft-state phase, a strong linear correlation between radio and X-ray fluxes is observed (Pearson $r$ = 0.97 for soft X-rays and $r$ = 0.98 for hard X-rays), while the disk luminosity remains nearly constant. At the outburst peak, a 5.14~mJy radio flux coincided with 2.6 (MAXI 2–10~keV) and 0.10 (BAT 15–50~keV) photons~cm$^{-2}$~s$^{-1}$, whereas the radio minimum (0.52~mJy) corresponded to 0.3 (2–20~keV) and 0.008 (15–50~keV) photons~cm$^{-2}$~s$^{-1}$. This tight correlation indicates that the radio emission traces contemporaneous jet activity directly powered by accretion processes.

The behavior contrasts with the 2020 outburst, where the first data point shows a very high radio flux followed by a rapid decline and no correlation with X-rays. These seem to be favorable to an inference that the first point could be in a main flux part of a transient jet but the following points could be in a long tail of the jet in time due to the jet-ISM interaction rather than central engine activity. As seen from Supplementary Figure~4, similar behavior was observed during the 2012 outburst of the source \cite{Neilsen2014}, where variable radio flux (0.25–0.75~mJy at 5.5~GHz) occurred despite constant hard X-ray flux, supporting an ejecta–ISM origin. Therefore, the 2020 outburst data are not appropriate for investigating X-ray properties when radio emission is detected. They are excluded in the analysis of relations between the radio flux and the semi-simultaneously obtained X-ray properties.

We also detected weak radio emissions in the 2022--2023 outburst on MJD = 60015 (second point of the radio light curve in the bottom left panel of Supplementary Figure~3). This observation coincides with the two highest points in the right panel, that is, the counts rate detected by MAXI and SWIFT/BAT is also the highest at this time; especially the hard X-ray counts rate is significantly higher than other observations. We are inclined to interpret them as indicative of the activity of the still-variable central source.

Supplementary Figure~5 summarizes the relationships among the fluxes of individual spectral components, radio flux densities, and the column densities of the X-ray winds. Overall, the disk flux remains consistent across different observations. While observations with stronger radio emission tend to be associated with higher coronal flux, the disk remains constant and continues to dominate the spectrum.

\subsection{Comparison with other near-Eddington sources.}

While sources like GRS~1915+105\cite{Done2004} and XTE~J1550--564\cite{Kubota2004} exhibit distinct soft-state regimes, probably associated with jet ejection events during regime transition; these occur at unusually high luminosities (Eddington ratios higher than 40\%, reaching super-Eddington in GRS~1915+105). These regimes likely reflect transitions between standard geometrically thin and slim accretion disks. In contrast, \src\ remains below $\sim$30\% of the Eddington limit throughout its outbursts, firmly in the standard disk regime without evidence for state transitions within the soft state. 

These distinct accretion regimes typically exhibit different PDS in GRS~1915+105 and XTE~J1550--564\cite{Reig2003, Inoue2022}. However, \src\ shows remarkably consistent timing properties with no significant variability power between both radio jet-active and X-ray wind-active states (Supplementary Figure~6). This consistency is further supported by spectral analysis using a disk model with a freely varying temperature profile exponent (p). The fits yield p=0.75 for both radio jet-active and X-ray wind-active states\cite{Kubota2004}. These results strongly favor interpretation within a standard accretion disk framework. We conclude that \src\ remains in a single, stable canonical soft state as in LMXBs, distinguished from high-Eddington sources that transition between accretion regimes.

\subsection{The launching radius and mass outflow rate of the X-ray winds.}

With the assumption that the outflow velocity of the winds is larger than the local escape velocity, we can estimate the lower limit of the radial location of the winds: $r_{\rm min} = 2GM_{\rm BH}v^{-2}$. During our monitoring program, the X-ray disk winds in \src\ remain $v\sim 1000$~km/s in 2020 and 2022-2023. By adopting this value, we estimate the radial location of the winds to be $r_{\rm min} \sim 10^{4}$~$R_{\rm g}$ ($10^{10}$\,cm).

We also calculate the upper limit of the launching location of winds by geometric consideration. As the thickness of the winds $\Delta R$ cannot exceed its distance to the central black hole $r$, we would have: $r_{\rm max}= L_{\rm ion}/N_{\rm H}\xi$. In this way, the maximum of the radial location will be $10^4 \sim 10^5$~$R_{\rm g}$ ($10^{10-11}$\,cm). 

Our estimations based on two methods suggest the X-ray winds in \src\ is likely launched at a radius similar to Compton radius $R_{\rm IC}\approx 10^{11-13}$\,cm assuming a black hole mass of 10\,M$_{\odot}$ and Compton temperature of $10^{6-8}$\,K as in a typical BH~XRB.

The mass outflow rate carried away by the X-ray winds can be evaluated as\cite{Ponti2012, Zhang2024, Neilsen2023} $$\Dot{M}_{\rm wind}=4\pi R^2nm_{\rm p}v_{\rm out}\frac{\Omega}{4\pi}=4\pi m_{\rm p}v_{\rm out}\frac{L_{\rm X}}{\xi}\frac{\Omega}{4\pi}$$ where $m_{\rm p}$ is the proton mass, $\Omega/4\pi$ is the covering factor, $v_{\rm out}$ is the wind outflow velocity. The covering factor, outflow velocity, luminosity $L_{\rm X}$ and ionization parameter $\xi$ all contribute to the uncertainties in this calculation\cite{Ponti2012}. We exploit a typical outflow velocity of $\sim 1000$~km/s. A black hole mass of 10.0~$M_{\bigodot}$ and a distance of 10~kpc are considered. With an assumed covering factor of $0.1-1.0$, we derive $\Dot{M}_{\rm wind} \sim 10^{17-18} \, \, {\rm g} \, \, {\rm s}^{-1}$. Similar mass outflow rates were measured for the previous outbursts of \src\ by Chandra and Suzaku \cite{Pahari2018, Gatuzz2019, Hori2014, Hori2018}. These wind properties are similar to those measured in other BH XRBs, such as H~1743--322\cite{Miller2006} and MAXI~J1803-298\cite{Zhang2024}.

\subsection{The estimation of the jets mass outflow rate.}

Quantifying jet properties requires associating observed characteristics with physical jet components to estimate minimum energy and mass outflow rates \cite{Longair1994, Fender1999}. While such calculations typically need proper motion measurements of jet components \cite{Fender1999, Fender2000} (unavailable in our data), we adopt parameters comparable to GRS~1915+105 \cite{Fender1999, Fender2000} as a reference. For our calculations for the radio jet-active soft state in 2021, we use the MeerKAT central frequency of 1.28 GHz and assume a distance of 10 kpc. The average observed flux density during this period at 1.28~GHz is 3~mJy, which would correspond to $\sim4$~mJy at 1~GHz, assuming the spectral index to be $-0.8$. Based on the assumption that proper motion is comparable to the case of GRS~1915+105\cite{Fender1999}, the observed emission at 1~GHz would have been emitted at 3~GHz, and its apparent flux density if observed in its rest frame would be about 80~mJy, leading to a luminosity of $L_{\rm \nu} \sim 10^{15} \, {\rm W} \, {\rm Hz}^{-1}$.

The mass outflow rate calculation carries significant uncertainties, primarily due to our limited constraints on jet event rise times ($\Delta t$) and source size. Following standard methodology\cite{Fender2000}, we estimate the emitting region volume as $V = (4/3)\pi(c\Delta t)^3$. For \src, we adopt rise times comparable to GRS~1915+105 observations: 12 hours ($V=10^{40}$~cm$^3$) and 20 minutes ($V=10^{33}$~cm$^3$)\cite{Fender1999,Fender2000}. Using these volume values, we derive mass outflow rates of $\sim 10^{16-19} \, \, {\rm g} \, \, {\rm s}^{-1}$ (assuming one proton per electron). While these estimates span several orders of magnitude, they remain consistent with jets in comparable systems\cite{Fender1999}.

\subsection{X-ray wind's column density verse radio flux density.} 

Our estimates show that the mass outflow rate carried by the X-ray disk winds in the 2022-2023 outburst is comparable to that carried by the jets in the 2021 outburst, and seems to be regulated by the law of conservation of mass.

Supplementary Figure~7 shows the X-ray wind’s column density versus radio flux density measured by pairs of X-ray and radio observations, in which we consider the simultaneous/quasi-simultaneous X-ray and radio observations with $\Delta$MJD$<$2. The inverse correlation between the X-ray wind’s column density and the radio flux density indicates the mutual exclusiveness between disk wind strength and jet intensity. Such a correlation is also implied in other black hole binary systems, like H1743-322\cite{Miller2006}. We run Spearman correlation analysis to quantify the anti-correlation between X-ray wind’s column density and radio flux density in Supplementary Figure~7. Our analysis suggests a correlation coefficient of $\rho = -0.47$ with significance of $>$~97\%. 

\subsection{Toward an unambiguous scenario for multi-wavelength outflows.}

Recent optical and infrared studies of black hole transients have substantially increased the number of XRBs with confirmed low-ionization (cold) disk winds, such as V404~Cyg\cite{Munoz2016} and MAXI~J1820+070\cite{Munoz2019}. These winds are inferred from distinctive spectroscopic features, including P-Cygni profiles, blue-shifted absorption troughs, and asymmetric broad emission line wings in optical and infrared wavelengths. Optical wind signatures are primarily observed during the hard state and sometimes simultaneously with the jets\cite{Munoz2019, Mata2022}. In the NIR, wind signatures have been reported in only a few systems, appearing in lines such as Pa$\beta$, Pa$\gamma$ and Br$\gamma$. Unlike their optical counterparts, NIR winds are observed in both hard and soft states, suggesting that wind-type ejecta can persist throughout the outburst cycle\cite{Sanchez2020, Sanchez2023}. 

It remains unclear whether X-ray and optical disk winds represent fundamentally different types of outflows or different manifestations of the same phenomenon\cite{Munoz2022, Zhang2024}. A unified picture is difficult to establish. X-ray disk winds are observed almost exclusively in systems with intermediate to high inclinations\cite{Ponti2012}, whereas optical and infrared winds are detected across a broader range of inclination angles, including low-inclination systems\cite{Ambrifi2025}. The presence of X-ray winds during the hard state has yet to be firmly established: they may be absent\cite{Ueda2010}, not detectable because of the highly ionized effect\cite{Shidatsu2019}, or suppressed by thermal instability\cite{Chakravorty2013}. Optical or infrared wind has not been reported for \src, which could otherwise provide valuable constraints. Achieving an unambiguous scenario for multi-wavelength outflows requires further high-quality observations and theoretical development.

\end{methods}

\begin{addendum}
 \item [Data availability] All data supporting the findings of this study are listed in the tables (Supplementary Tables 1, 2, and 3)  within the paper. The MeerKAT and NICER data used in this study are publicly available, and corresponding observation IDs are listed in the tables within the paper. The MeerKAT observations were largely performed as part of the ThunderKAT, which is a large MeerKAT open-time programme to observe X-ray binaries in the radio band. The data can be accessed from MeerKAT science archive (https://archive.sarao.ac.za/). The NICER observations can be accessed from the HEASARC science archive (https://heasarc.gsfc.nasa.gov/docs/archive.html).
 \item [Acknowledgements] This work was supported by the National Natural Science Foundation of China (NSFC), Grant No.~12250610185, 11973019, and 12261131497, and the Natural Science Foundation of Shanghai, Grant No. 22ZR1403400. JvdE acknowledges a Warwick Astrophysics prize post-doctoral fellowship made possible thanks to a generous philanthropic donation, and was supported by funding from the European Union's Horizon Europe research and innovation programme under the Marie Skłodowska-Curie grant agreement No 101148693 (MeerSHOCKS) for part of this work.  
 \item [Author Contributions Statement] Z.Z. led the analysis and interpretation of X-ray data, and drafted the manuscript. J.J. and C.B. contributed to the X-ray data analysis and interpretation. F.C. carried out and led the analysis and interpretation of the radio observations. All authors discussed the results and contributed to the final manuscript.
 \item[Competing Interests Statement] The authors declare that they have no competing interests.
 
\end{addendum}



\clearpage

\section*{References}


\newcommand{\actaa}{Acta Astron.}   
\newcommand{\araa}{Annu. Rev. Astron. Astrophys.}   
\newcommand{\areps}{Annu. Rev. Earth Planet. Sci.} 
\newcommand{\aar}{Astron. Astrophys. Rev.} 
\newcommand{\ab}{Astrobiology}    
\newcommand{\aj}{Astron. J.}   
\newcommand{\ac}{Astron. Comput.} 
\newcommand{\apart}{Astropart. Phys.} 
\newcommand{\apj}{Astrophys. J.}   
\newcommand{\apjl}{Astrophys. J. Lett.}   
\newcommand{\apjs}{Astrophys. J. Suppl. Ser.}   
\newcommand{\ao}{Appl. Opt.}   
\newcommand{\apss}{Astrophys. Space Sci.}   
\newcommand{\aap}{Astron. Astrophys.}   
\newcommand{\aapr}{Astron. Astrophys. Rev.}   
\newcommand{\aaps}{Astron. Astrophys. Suppl.}   
\newcommand{\baas}{Bull. Am. Astron. Soc.}   
\newcommand{\caa}{Chin. Astron. Astrophys.}   
\newcommand{\cjaa}{Chin. J. Astron. Astrophys.}   
\newcommand{\cqg}{Class. Quantum Gravity}    
\newcommand{\epsl}{Earth Planet. Sci. Lett.}    
\newcommand{\expa}{Exp. Astron.}    
\newcommand{\frass}{Front. Astron. Space Sci.}    
\newcommand{\gal}{Galaxies}    
\newcommand{\gca}{Geochim. Cosmochim. Acta}   
\newcommand{\grl}{Geophys. Res. Lett.}   
\newcommand{\icarus}{Icarus}   
\newcommand{\ija}{Int. J. Astrobiol.} 
\newcommand{\jatis}{J. Astron. Telesc. Instrum. Syst.}  
\newcommand{\jcap}{J. Cosmol. Astropart. Phys.}   
\newcommand{\jgr}{J. Geophys. Res.}   
\newcommand{\jgrp}{J. Geophys. Res.: Planets}    
\newcommand{\jheap}{J. High Energy Astrophys.}    
\newcommand{\joss}{J. Open Source Softw.}    
\newcommand{\jqsrt}{J. Quant. Spectrosc. Radiat. Transf.} 
\newcommand{\lrca}{Living Rev. Comput. Astrophys.}    
\newcommand{\lrr}{Living Rev. Relativ.}    
\newcommand{\lrsp}{Living Rev. Sol. Phys.}    
\newcommand{\memsai}{Mem. Soc. Astron. Italiana}   
\newcommand{\maps}{Meteorit. Planet. Sci.} 
\newcommand{\mnras}{Mon. Not. R. Astron. Soc.}   
\newcommand{\nat}{Nature} 
\newcommand{\nastro}{Nat. Astron.} 
\newcommand{\ncomms}{Nat. Commun.} 
\newcommand{\ngeo}{Nat. Geosci.} 
\newcommand{\nphys}{Nat. Phys.} 
\newcommand{\na}{New Astron.}   
\newcommand{\nar}{New Astron. Rev.}   
\newcommand{\physrep}{Phys. Rep.}   
\newcommand{\pra}{Phys. Rev. A}   
\newcommand{\prb}{Phys. Rev. B}   
\newcommand{\prc}{Phys. Rev. C}   
\newcommand{\prd}{Phys. Rev. D}   
\newcommand{\pre}{Phys. Rev. E}   
\newcommand{\prl}{Phys. Rev. Lett.}   
\newcommand{\psj}{Planet. Sci. J.}   
\newcommand{\planss}{Planet. Space Sci.}   
\newcommand{\pnas}{Proc. Natl Acad. Sci. USA}   
\newcommand{\procspie}{Proc. SPIE}   
\newcommand{\pasa}{Publ. Astron. Soc. Aust.}   
\newcommand{\pasj}{Publ. Astron. Soc. Jpn}   
\newcommand{\pasp}{Publ. Astron. Soc. Pac.}   
\newcommand{\raa}{Res. Astron. Astrophys.} 
\newcommand{\rasti}{RAS Tech. Instrum.} 
\newcommand{\rmxaa}{Rev. Mexicana Astron. Astrofis.}   
\newcommand{\rnaas}{Res. Notes Am. Astron. Soc.} 
\newcommand{\sci}{Science} 
\newcommand{\sciadv}{Sci. Adv.} 
\newcommand{\solphys}{Sol. Phys.}   
\newcommand{\sovast}{Soviet Astron.}   
\newcommand{\ssr}{Space Sci. Rev.}   
\newcommand{\uni}{Universe} 




\newpage

\renewcommand{\thefigure}{\arabic{figure}}
\renewcommand{\thetable}{\arabic{table}}
\setcounter{figure}{0}
\setcounter{table}{0}

\makeatletter
\renewcommand{\fps@figure}{H}   
\renewcommand{\fps@table}{H}    
\makeatother

\captionsetup[table]{labelformat=empty, font=footnotesize, labelfont=bf, labelsep=period}
\captionsetup[figure]{labelformat=empty, font=footnotesize, labelfont=bf, labelsep=period}


\fontsize{7}{7}\selectfont
\setlength{\tabcolsep}{2.4pt}
\renewcommand{\arraystretch}{1.1}

\begin{longtable}{lccccc}
\caption{\textbf{Supplementary Table \thetable~\boldmath{$|$} Overview of the NICER observations used in this analysis.} MJD\_s and MJD\_e indicate the time when the observation starts and ends, respectively. The count rate is for 52 active detectors.}  \label{nicer_obser}  \\

\toprule
Num   & Obs ID    & MJD\_s(days)   & MJD\_e(days)   & Rate(cts/s)  &  Exposure(s)       \\ 
\midrule
\endfirsthead

\toprule
Num   & Obs ID    & MJD\_s(days)   & MJD\_e(days)   & Rate(cts/s)  &  Exposure(s)       \\ 
\midrule
\endhead	

\bottomrule
\multicolumn{3}{l}{\textit{Continued on next page...}} \\
\endfoot
\bottomrule
\endlastfoot 

    1 & 3130010101   & 58927.657 & 58927.663 & 395.11 & 565.80 \\
    2 & 3130010102   & 58934.673 & 58934.682 & 561.86 & 757.80 \\
    3 & 3130010103   & 58935.900 & 58935.911 & 621.51 & 956.80 \\
    4 & 3130010104   & 58936.416 & 58936.427 & 613.18 & 977.80 \\
    5 & 3130010105   & 58938.223 & 58938.227 & 611.79 & 299.80 \\
    6 & 3130010106   & 58940.870 & 58940.882 & 612.88 & 1009.80 \\
    7 & 3130010107   & 58941.645 & 58941.716 & 574.04 & 1145.60 \\
    8 & 3130010108   & 58942.613 & 58942.625 & 512.02 & 1031.80 \\
    9 & 3130010109   & 58943.130 & 58943.529 & 503.23 & 1087.60 \\
   10 & 3130010110   & 58945.018 & 58945.027 & 525.82 & 749.60 \\
   11 & 3130010111   & 58946.052 & 58946.564 & 501.01 & 2542.40 \\
   12 & 3130010112   & 58948.700 & 58948.837 & 528.08 & 1965.40 \\
   13 & 3130010114   & 58950.556 & 58950.561 & 522.45 & 388.80 \\
   14 & 3130010116   & 58952.238 & 58952.372 & 539.38 & 919.40 \\
   15 & 3130010117   & 58953.144 & 58953.278 & 583.16 & 947.40 \\
   16 & 3130010118   & 58954.821 & 58954.890 & 544.38 & 953.40 \\
   17 & 3130010119   & 58956.107 & 58956.112 & 634.09 & 393.80 \\
   18 & 3130010120   & 58957.804 & 58957.806 & 641.49 & 110.80 \\
   19 & 3130010121   & 58958.194 & 58958.266 & 685.19 & 1327.60 \\
   20 & 3130010123   & 58960.432 & 58960.439 & 662.38 & 608.80 \\
   21 & 3130010124   & 58961.160 & 58961.167 & 705.71 & 611.60 \\
   22 & 3130010125   & 58962.261 & 58962.267 & 696.53 & 513.60 \\
   23 & 3130010126   & 58963.013 & 58963.476 & 732.46 & 1011.20 \\
   24 & 3130010127   & 58965.014 & 58965.736 & 776.11 & 1485.60 \\
   25 & 3130010128   & 58966.562 & 58966.638 & 700.33 & 1544.60 \\
   26 & 3130010129   & 58967.272 & 58967.920 & 677.52 & 1286.60 \\
   27 & 3130010130   & 58968.179 & 58968.373 & 704.80 & 1813.20 \\
   28 & 3130010131   & 58969.080 & 58969.286 & 697.92 & 3132.20 \\
   29 & 3130010132   & 58970.115 & 58970.190 & 605.84 & 1385.60 \\
   30 & 3130010133   & 58971.342 & 58971.417 & 560.49 & 1745.60 \\
   31 & 3130010134   & 58972.310 & 58972.966 & 576.68 & 1831.20 \\
   32 & 3130010135   & 58973.020 & 58973.353 & 596.90 & 1042.60 \\
   33 & 3130010136   & 58974.311 & 58974.450 & 570.29 & 2580.20 \\
   34 & 3130010137   & 58975.409 & 58975.482 & 555.87 & 1529.80 \\
   35 & 3130010138   & 58976.378 & 58976.902 & 510.34 & 1355.20 \\
   36 & 3130010139   & 58977.412 & 58977.483 & 520.87 & 1174.40 \\
   37 & 3130010140   & 58978.188 & 58978.451 & 517.35 & 1965.80 \\
   38 & 3130010144   & 58991.505 & 58991.707 & 253.41 & 2903.00 \\
   39 & 3130010145   & 58997.961 & 58997.976 & 158.88 & 1221.80 \\
   40 & 3130010146   & 59000.091 & 59000.107 & 144.37 & 743.40 \\
   41 & 3130010147   & 59001.774 & 59001.848 & 135.13 & 1625.60 \\
   42 & 4130010101   & 59470.787 & 59470.926 & 42.98 & 1610.00 \\
   43 & 4130010102   & 59471.110 & 59471.831 & 53.67 & 4675.20 \\
   44 & 4130010103   & 59472.087 & 59472.808 & 62.83 & 3943.20 \\
   45 & 4130010104   & 59473.063 & 59473.974 & 73.84 & 3499.00 \\
   46 & 4130010105   & 59474.034 & 59474.933 & 96.10 & 3581.40 \\
   47 & 4130010106   & 59475.243 & 59475.588 & 135.95 & 2613.00 \\
   48 & 4130010107   & 59476.084 & 59476.928 & 164.78 & 4670.00 \\
   49 & 4130010108   & 59477.179 & 59477.961 & 192.07 & 3378.00 \\
   50 & 4130010110   & 59479.698 & 59479.705 & 326.95 & 600.80 \\
   51 & 4130010111   & 59480.537 & 59480.933 & 376.62 & 2633.00 \\
   52 & 4130010112   & 59481.054 & 59481.910 & 237.88 & 2809.00 \\
   53 & 4130010113   & 59482.806 & 59482.816 & 252.09 & 889.80 \\
   54 & 4130010114   & 59483.258 & 59483.654 & 333.66 & 1809.40 \\
   55 & 4130010115   & 59484.419 & 59484.879 & 691.35 & 1456.40 \\
   56 & 4130010116   & 59485.319 & 59485.909 & 969.15 & 1063.20 \\
   57 & 4130010118   & 59486.997 & 59487.003 & 981.39 & 534.80 \\
   58 & 4130010121   & 59498.005 & 59498.659 & 756.46 & 1565.40 \\
   59 & 4130010122   & 59499.168 & 59499.880 & 744.91 & 785.60 \\
   60 & 4130010123   & 59500.203 & 59500.211 & 658.77 & 684.60 \\
   61 & 4130010124   & 59501.239 & 59501.698 & 674.58 & 1164.40 \\
   62 & 4130010125   & 59502.022 & 59502.540 & 579.72 & 773.00 \\
   63 & 4130010126   & 59508.906 & 59508.913 & 602.34 & 604.80 \\
   64 & 4130010127   & 59510.455 & 59510.464 & 456.51 & 777.80 \\
   65 & 4130010128   & 59512.462 & 59512.468 & 517.20 & 411.40 \\
   66 & 6130010107   & 60013.049 & 60013.060 & 919.28 & 944.80 \\
   67 & 6557010102   & 60014.018 & 60014.163 & 946.68 & 3730.80 \\
   68 & 6557010302   & 60016.022 & 60016.811 & 1199.78 & 9203.60 \\
   69 & 6130010109   & 60030.336 & 60030.347 & 814.57 & 919.60 \\
   70 & 5665010406   & 60039.886 & 60039.898 & 916.88 & 667.60 \\
   71 & 5665010407   & 60041.371 & 60041.639 & 789.52 & 3582.20 \\
   72 & 5665010408   & 60042.661 & 60042.670 & 714.52 & 778.80 \\
   73 & 6130010110   & 60050.805 & 60050.874 & 753.60 & 726.60 \\
   74 & 5665010409   & 60055.196 & 60055.269 & 661.77 & 1309.40 \\
   75 & 6130010111   & 60056.871 & 60056.946 & 568.63 & 1739.60 \\
   76 & 6130010112   & 60057.000 & 60057.075 & 562.15 & 1711.60 \\
   77 & 6130010113   & 60065.371 & 60065.831 & 529.30 & 2899.80 \\
   78 & 5665010410   & 60069.309 & 60069.508 & 648.02 & 1934.20 \\
   79 & 5665010411   & 60083.524 & 60083.661 & 492.46 & 1841.00 \\
   80 & 6130010117   & 60085.464 & 60085.601 & 528.44 & 1605.00 \\
   81 & 6130010118   & 60092.815 & 60092.895 & 659.96 & 2673.00 \\
   82 & 6130010119   & 60103.192 & 60103.458 & 419.44 & 2196.40 \\
   83 & 6130010120   & 60111.515 & 60111.585 & 449.71 & 938.60 \\
   84 & 6130010121   & 60113.459 & 60113.842 & 403.63 & 2029.20 \\

\end{longtable}


\newpage


\begin{longtable}{lccccccc}
\caption{\textbf{Supplementary Table \thetable~\boldmath{$|$} Overview of the MeerKAT observations used in this analysis.} State indicates the accretion state of the source during the observation period as obtained from X-ray data analysis. Errors represent 1$\sigma$ confidence intervals, while 3$\sigma$ upper limits are provided for non-detections.}  \label{meerkat_obser}  \\

\toprule
Num & Obs ID  & MJD(days)        & Flux(mJy)  & Error(mJy)   & Spectral index    & Error & State \\ 
\midrule
\endfirsthead

\toprule
Num & Obs ID  & MJD(days)        & Flux(mJy)  & Error(mJy)   & Spectral index    & Error & State \\
\midrule
\endhead	

\bottomrule
\multicolumn{3}{l}{\textit{Continued on next page...}} \\
\endfoot
\bottomrule
\endlastfoot 

   1 & 1584753662 & 58929.085          & 4.77     &   0.08      &  $-0.7$  &  0.4  &    Soft         \\
   2 & 1585466272 & 58937.333          & 0.9       &   0.2        &   -      &    -    &    Soft         \\
   3 & 1585883785 &58942.151          & 0.54     &   0.03      &   -      &    -    &    Soft         \\
   4 & 1586498460 & 58949.265          & 0.64     &   0.03      &   -      &    -    &    Soft         \\
   5 & 1587176616 & 58957.114          & 0.30     &   0.04      &   -      &    -    &    Soft         \\
   6 & 1587786356 & 58964.171         & $ < $0.6   &    -          &   -      &    -    &    Soft         \\  
   7 & 1588375604 & 58970.991         & $ < $0.3   &    -          &   -      &    -    &    Soft         \\               
   8 & 1588992355 & 58978.144         & $ < $0.45   &    -        &   -      &    -    &    Soft         \\               
   9 & 1589683557 & 58986.144         & $ < $0.6   &    -          &   -      &    -    &    Soft         \\  
 10 & 1593890159 & 59034.808         & $ < $0.3   &    -          &   -      &    -    &    Soft         \\          
 11 & 1594580523 & 59042.799         &  0.35      &   0.02     &   -      &    -    &    Soft         \\  
 12 & 1595185558 & 59049.801        &  $ < $0.45   &    -        &   -      &    -    &    Soft         \\  
 13 & 1595623962 & 59054.690        &  $ < $0.3   &    -          &   -      &    -   &    Soft         \\  
 14 & 1632148282 & 59477.639        &   1.77     &   0.05      &  $-0.1$  &  0.4  &    Hard         \\
 15 & 1632745869 & 59484.385        &   5.14     &   0.04      &  $-0.7$  &  0.4  &    Soft         \\       
 16 & 1633356072 & 59491.616        &   1.84     &   0.03      &  $-1.0$  &  0.4  &    Soft         \\      
 17 & 1633788074 & 59496.616        &   2.58     &   0.05      &  $-1.4$  &  0.5  &    Soft         \\ 
 18 & 1634467571 & 59504.48         &    1.90     &   0.04      &  $-1.2$  &  0.5  &    Soft         \\   
 19 & 1634986873 & 59510.49         &   2.18     &   0.04      &  $-0.5$  &  0.5  &    Soft         \\   
 20 & 1635685270 & 59518.574        &   1.77     &   0.08      &  $-0.5$   &  0.5  &    Soft         \\ 
 21 & 1636886466 & 59532.477        &   0.96     &   0.03      &   -      &    -    &    Soft         \\
 22 & 1637335706 & 59537.651        &   0.94     &   0.05      &   -      &    -    &    Soft         \\
 23 & 1638001320 & 59545.369        &   0.52     &   0.04      &   -      &    -    &    Soft         \\
 24 & 1645173077 & 59628.385        &   $ < $0.45   &    -          &   -      &    -    &    Soft         \\ 
 25 & 1678494193 & 60014.022        &   $ < $0.45   &    -          &   -      &    -    &    Soft         \\               
 26 & 1678597938 & 60015.229        &   0.26     &   0.04      &   -      &    -    &    Soft         \\
 27 & 1679703078 & 60028.021        &   $ < $0.45   &    -          &   -      &    -    &    Soft         \\ 
 28 & 1680330672 & 60035.285        &   $ < $0.6   &    -           &   -      &    -    &    Soft         \\     
 29 & 1680927674 & 60043.194        &   $ < $0.45   &    -          &   -      &    -    &    Soft         \\ 
 30 & 1682135479 & 60056.17         &    $ < $0.45   &    -          &   -      &    -    &    Soft         \\ 
 31 & 1682819781 & 60064.17         &    $ < $0.3   &    -          &   -      &    -    &    Soft         \\ 
 32 & 1683339854 & 60070.11         &    $ < $0.45   &    -          &   -      &    -    &    Soft         \\ 
 33 & 1684103257 & 60078.94         &    $ < $0.3   &    -          &   -      &    -    &    Soft         \\ 
 34 & 1684711875 & 60085.99         &    $ < $0.3   &    -          &   -      &    -    &    Soft         \\ 
 35 & 1685152874 & 60091.008         &    $ < $0.3   &    -          &   -      &    -    &    Soft         \\ 
 36 & 1685742346 & 60097.92         &     $ < $0.4   &    -          &   -      &    -    &    Soft         \\ 
 37 & 1686520945 & 60106.92         &     $ < $0.45   &    -          &   -      &    -    &    Soft         \\ 
 38 & 1686966372 & 60112.008         &     $ < $0.45   &    -          &   -      &    -    &    Soft         \\ 
\end{longtable}


\newpage



\begin{longtable}{lcccccccccc}
\caption{\textbf{Supplementary Table \thetable~\boldmath{$|$} Results of the spectral fitting to measure the X-ray wind's strength.} The details are described in Methods section. Error bars represent 3$\sigma$ confidence intervals. Absorption records whether or not a significant X-ray disk wind features was observed. The fluxes are in units of erg~cm$^{-2}$~s$^{-1}$. Parameters with * are fixed during the fit.}  \label{fit_results}  \\

\toprule
Obs ID   & State  &  Tbabs                                 &  \multicolumn{3}{c}{XSTAR}                                                  & \multicolumn{2}{c}{diskbb}                      &  \multicolumn{2}{c}{nthcomp}    \\ 
           &              & N$_{\rm H}$ ($10^{22}$~cm$^{-2}$)  &  N$_{\rm H}$ ($10^{22}$~cm$^{-2}$)               & $\log(\xi)$ &  $z$  & kT$_{\rm in}$ (keV) & $\log F_{\rm diskbb}$   &  $\Gamma$  & $\log F_{\rm nthcomp}$     \\
\midrule
\endfirsthead

\toprule
Obs ID   & State  &  Tbabs                                 &  \multicolumn{3}{c}{XSTAR}                                                  & \multicolumn{2}{c}{diskbb}                      &  \multicolumn{2}{c}{nthcomp}    \\ 
           &              & N$_{\rm H}$ ($10^{22}$~cm$^{-2}$)   &  N$_{\rm H}$ ($10^{22}$~cm$^{-2}$)                  & $\log(\xi)$ &  $z$  & kT$_{\rm in}$ (keV)  & $\log F_{\rm diskbb}$   &  $\Gamma$  & $\log F_{\rm nthcomp}$     \\
\midrule
\endhead	

\bottomrule
\multicolumn{3}{l}{\textit{Continued on next page...}} \\
\endfoot
\bottomrule
\endlastfoot 

3130010101 & Soft & ${11.8}^{*}$                   & $<1.5$          & ${4.2}^{*}$                    & ${-0.004}^{*}$                 & $1.233_{-0.022}^{+0.021}$      & $-8.033_{-0.007}^{+0.007}$     & ${2.0}^{*}$                    & $-8.68_{-0.04}^{+0.04}$      \\
3130010102 & Soft & $11.85_{-0.11}^{+0.11}$        & $<3.8$                & ${4.4}^{*}$                    & ${-0.004}^{*}$                 & $1.40_{-0.03}^{+0.03}$         & $-7.822_{-0.005}^{+0.005}$     & ${2.0}^{*}$                    & $-9.00_{-0.13}^{+0.10}$        \\
3130010103 & Soft & $12.00_{-0.09}^{+0.09}$        & $<2.6$              & ${4.4}^{*}$                    & ${-0.004}^{*}$                 & $1.428_{-0.027}^{+0.029}$      & $-7.775_{-0.004}^{+0.004}$     & ${2.0}^{*}$                    & $-8.93_{-0.11}^{+0.09}$       \\
3130010104 & Soft & $11.92_{-0.09}^{+0.09}$        & $<1.2$            & ${4.4}^{*}$                    & ${-0.004}^{*}$                 & $1.402_{-0.027}^{+0.028}$      & $-7.787_{-0.004}^{+0.004}$     & ${2.0}^{*}$                    & $-8.90_{-0.10}^{+0.08}$       \\ 
3130010105 & Soft & $11.99_{-0.16}^{+0.16}$        & $<5.6$           & ${4.4}^{*}$                    & ${-0.004}^{*}$                 & $1.46_{-0.05}^{+0.05}$         & $-7.769_{-0.007}^{+0.007}$     & ${2.0}^{*}$                    & $-9.27_{-0.8}^{+0.26}$      \\  
3130010106 & Soft & $12.00_{-0.09}^{+0.09}$        & $<4.5$           & ${4.4}^{*}$            & ${-0.004}^{*}$      & $1.457_{-0.026}^{+0.027}$      & $-7.764_{-0.004}^{+0.004}$     & ${2.0}^{*}$                    & $-9.4_{-0.4}^{+0.20}$      \\    
3130010107 & Soft & $12.01_{-0.09}^{+0.09}$        & $4.1_{-2.9}^{+5}$             & $4.3_{-0.4}^{+0.7}$            & $>-0.012$     & $1.415_{-0.023}^{+0.024}$      & $-7.790_{-0.004}^{+0.003}$     & ${2.0}^{*}$                    & $-9.37_{-0.28}^{+0.17}$      \\ 
3130010108 & Soft & $11.89_{-0.09}^{+0.09}$        & $7_{-6}^{+10}$            & $>3.9$            & $>-0.005$     & $1.409_{-0.025}^{+0.022}$      & $-7.837_{-0.004}^{+0.004}$     & ${2.0}^{*}$                    & $<-9.5$            \\
3130010109 & Soft & $11.83_{-0.08}^{+0.09}$        & $20.6_{-9}^{+1.3}$            & $5.7_{-1.3}^{+1.1}$          & $>-0.002$     & $1.422_{-0.016}^{+0.009}$      & $-7.8422_{-0.004}^{+0.0022}$   & ${2.0}^{*}$                    & $<-9.6$     \\        
3130010110 & Soft & $12.04_{-0.09}^{+0.08}$        & $20.6_{-14}^{+1.4}$            & $>3.9$          & $>-0.008$     & $1.430_{-0.012}^{+0.011}$      & $-7.821_{-0.004}^{+0.004}$     & ${2.0}^{*}$                    & $<-9.8$     \\    
3130010111 & Soft & $11.86_{-0.05}^{+0.04}$        & $7.7_{-1.8}^{+6}$            & $4.36_{-0.12}^{+0.7}$         & $>-0.004$  & $1.435_{-0.011}^{+0.006}$      & $-7.8462_{-0.0023}^{+0.0022}$  & ${2.0}^{*}$                    & $<-9.7$    \\     
3130010112 & Soft & $11.89_{-0.04}^{+0.06}$        & $12.7_{-2.6}^{+7}$            & $4.72_{-0.18}^{+1.8}$         & $-0.0025_{-0.0020}^{+0.0020}$   & $1.457_{-0.017}^{+0.010}$       & $-7.8220_{-0.0027}^{+0.0025}$  & ${2.0}^{*}$                    & $<-9.8$       \\    
3130010114 & Soft & $11.99_{-0.11}^{+0.11}$        & $13_{-9}^{+9}$            & $>3.2$              & $>-0.022$     & $1.420_{-0.026}^{+0.015}$      & $-7.819_{-0.006}^{+0.005}$     & ${2.0}^{*}$                    & $<-9.7$        \\   
3130010116 & Soft & $12.05_{-0.10}^{+0.10}$           & $13_{-7}^{+7}$            & $>4.0$         & $>-0.003$     & $1.455_{-0.028}^{+0.018}$      & $-7.811_{-0.004}^{+0.004}$     & ${2.0}^{*}$                    & $<-9.5$      \\     
3130010117 & Soft & $12.10_{-0.07}^{+0.08}$        & $13_{-10}^{+4}$              & $>4.3$              & $>-0.008$        & $1.488_{-0.021}^{+0.009}$      & $-7.774_{-0.003}^{+0.003}$     & ${2.0}^{*}$                    & $<-9.8$       \\      
3130010118 & Soft & $11.910_{-0.10}^{+0.10}$           & $13_{-5}^{+10}$            & $>3.9$          & $>-0.006$     & $1.482_{-0.029}^{+0.029}$      & $-7.824_{-0.005}^{+0.004}$     & ${2.0}^{*}$                    & $<-9.4$    \\        
3130010119 & Soft & $11.99_{-0.09}^{+0.15}$        & $<5.0$             & ${4.4}^{*}$                    & ${-0.004}^{*}$                 & $1.563_{-0.05}^{+0.016}$       & $-7.7553_{-0.008}^{+0.0026}$   & ${2.0}^{*}$                    & $<-9.3$     \\      
3130010120 & Soft & $12.03_{-0.18}^{+0.19}$        & $<4.9$              & ${4.5}^{*}$                    & ${-0.004}^{*}$                 & $1.505_{-0.04}^{+0.023}$       & $-7.746_{-0.008}^{+0.003}$     & ${2.0}^{*}$                    & $<-9.4$       \\    
3130010121 & Soft & $12.08_{-0.07}^{+0.07}$        & $<3.8$           & ${4.5}^{*}$                    & ${-0.004}^{*}$                 & $1.475_{-0.023}^{+0.024}$      & $-7.723_{-0.003}^{+0.003}$     & ${2.0}^{*}$                    & $-9.02_{-0.13}^{+0.10}$      \\  
3130010123 & Soft & $11.76_{-0.11}^{+0.11}$        & $<1.8$              & ${4.4}^{*}$                    & ${-0.004}^{*}$                 & $1.41_{-0.04}^{+0.04}$         & $-7.766_{-0.005}^{+0.005}$     & ${2.0}^{*}$                    & $-8.81_{-0.11}^{+0.09}$       \\
3130010124 & Soft & $11.93_{-0.11}^{+0.11}$        & $7_{-5}^{+9}$              & $>4.2$            & $>-0.008$     & $1.50_{-0.04}^{+0.04}$         & $-7.717_{-0.005}^{+0.005}$     & ${2.0}^{*}$                    & $-8.96_{-0.19}^{+0.13}$      \\ 
3130010125 & Soft & $11.89_{-0.12}^{+0.12}$        & $<6.0$             & ${4.5}^{*}$                    & ${-0.004}^{*}$                 & $1.45_{-0.04}^{+0.04}$         & $-7.731_{-0.005}^{+0.005}$     & ${2.0}^{*}$                    & $-8.87_{-0.15}^{+0.11}$     \\  
3130010126 & Soft & $11.98_{-0.09}^{+0.09}$        & $<4.3$              & ${4.5}^{*}$                    & ${-0.004}^{*}$                 & $1.49_{-0.03}^{+0.03}$         & $-7.704_{-0.004}^{+0.004}$     & ${2.0}^{*}$                    & $-8.95_{-0.16}^{+0.11}$     \\  
3130010127 & Soft & $11.96_{-0.07}^{+0.07}$        & $<2.9$              & ${4.5}^{*}$                    & ${-0.004}^{*}$                 & $1.502_{-0.027}^{+0.028}$      & $-7.688_{-0.004}^{+0.004}$     & ${2.0}^{*}$                    & $-8.85_{-0.11}^{+0.09}$      \\ 
3130010128 & Soft & $11.89_{-0.07}^{+0.07}$        & $<5.4$                & ${4.5}^{*}$                    & ${-0.004}^{*}$                 & $1.448_{-0.025}^{+0.025}$      & $-7.735_{-0.004}^{+0.004}$     & ${2.0}^{*}$                    & $-8.81_{-0.08}^{+0.07}$       \\
3130010129 & Soft & $11.84_{-0.08}^{+0.08}$        & $<2.2$            & ${4.5}^{*}$                    & ${-0.004}^{*}$                 & $1.468_{-0.025}^{+0.026}$      & $-7.740_{-0.004}^{+0.004}$     & ${2.0}^{*}$                    & $-9.08_{-0.16}^{+0.11}$      \\ 
3130010130 & Soft & $11.82_{-0.06}^{+0.06}$        & $<1.3$            & ${4.5}^{*}$                    & ${-0.004}^{*}$                 & $1.468_{-0.021}^{+0.021}$      & $-7.7269_{-0.0029}^{+0.0029}$  & ${2.0}^{*}$                    & $-8.94_{-0.09}^{+0.07}$       \\
3130010131 & Soft & $11.77_{-0.05}^{+0.05}$        & $<1.5$              & ${4.5}^{*}$                    & ${-0.004}^{*}$                 & $1.484_{-0.016}^{+0.017}$      & $-7.7293_{-0.0023}^{+0.0022}$  & ${2.0}^{*}$                    & $-9.04_{-0.09}^{+0.07}$       \\
3130010132 & Soft & $11.59_{-0.08}^{+0.08}$        & $<1.3$              & ${4.4}^{*}$                    & ${-0.004}^{*}$                 & $1.337_{-0.023}^{+0.024}$      & $-7.825_{-0.004}^{+0.004}$     & ${2.0}^{*}$                    & $-8.68_{-0.05}^{+0.04}$      \\ 
3130010133 & Soft & $11.38_{-0.08}^{+0.07}$        & $<1.0$              & ${4.3}^{*}$                    & ${-0.004}^{*}$                 & $1.319_{-0.021}^{+0.022}$      & $-7.876_{-0.003}^{+0.003}$     & ${2.0}^{*}$                    & $-8.65_{-0.03}^{+0.03}$       \\
3130010134 & Soft & $11.52_{-0.07}^{+0.07}$        & $<0.8$              & ${4.4}^{*}$                    & ${-0.004}^{*}$                 & $1.358_{-0.020}^{+0.021}$       & $-7.839_{-0.003}^{+0.003}$     & ${2.0}^{*}$                    & $-8.77_{-0.05}^{+0.04}$     \\  
3130010135 & Soft & $11.62_{-0.09}^{+0.09}$        & $<1.2$            & ${4.4}^{*}$                    & ${-0.004}^{*}$                 & $1.401_{-0.027}^{+0.028}$      & $-7.811_{-0.004}^{+0.004}$     & ${2.0}^{*}$                    & $-8.90_{-0.09}^{+0.07}$      \\ 
3130010136 & Soft & $11.65_{-0.06}^{+0.06}$        & $<0.7$            & ${4.4}^{*}$                    & ${-0.004}^{*}$                 & $1.348_{-0.016}^{+0.017}$      & $-7.8338_{-0.0025}^{+0.0025}$  & ${2.0}^{*}$                    & $-8.82_{-0.04}^{+0.04}$       \\
3130010137 & Soft & $11.56_{-0.08}^{+0.08}$        & $<0.8$            & ${4.4}^{*}$                    & ${-0.004}^{*}$                 & $1.378_{-0.022}^{+0.022}$      & $-7.839_{-0.003}^{+0.003}$     & ${2.0}^{*}$                    & $-8.96_{-0.08}^{+0.07}$       \\
3130010138 & Soft & $11.54_{-0.09}^{+0.09}$        & $<2.3$              & ${4.3}^{*}$                    & ${-0.004}^{*}$                 & $1.318_{-0.023}^{+0.023}$      & $-7.889_{-0.004}^{+0.004}$     & ${2.0}^{*}$                    & $-8.84_{-0.06}^{+0.05}$       \\
3130010139 & Soft & $11.63_{-0.09}^{+0.04}$        & $<1.1$            & ${4.3}^{*}$                    & ${-0.004}^{*}$                 & $1.339_{-0.021}^{+0.022}$      & $-7.8670_{-0.0014}^{+0.0022}$  & ${2.0}^{*}$                    & $-8.930_{-0.024}^{+0.025}$    \\
3130010140 & Soft & $11.60_{-0.07}^{+0.07}$        & $<0.7$            & ${4.3}^{*}$                    & ${-0.004}^{*}$                 & $1.332_{-0.018}^{+0.019}$      & $-7.8698_{-0.003}^{+0.0029}$   & ${2.0}^{*}$                    & $-8.92_{-0.06}^{+0.05}$       \\
3130010144 & Soft & $11.53_{-0.09}^{+0.09}$        & $<1.5$              & ${4.0}^{*}$                    & ${-0.004}^{*}$                 & $1.120_{-0.013}^{+0.014}$      & $-8.156_{-0.005}^{+0.005}$     & ${2.0}^{*}$                    & $-9.32_{-0.05}^{+0.04}$     \\  
3130010145 & Soft & $11.33_{-0.19}^{+0.19}$        & $<2.9$              & ${3.8}^{*}$                    & ${-0.004}^{*}$                 & $0.987_{-0.024}^{+0.025}$      & $-8.370_{-0.012}^{+0.013}$     & ${2.0}^{*}$                    & $-9.24_{-0.05}^{+0.04}$       \\
3130010146 & Soft & $11.17_{-0.26}^{+0.27}$        & $<2.2$          & ${3.8}^{*}$                    & ${-0.004}^{*}$                 & $0.96_{-0.03}^{+0.03}$         & $-8.426_{-0.017}^{+0.019}$     & ${2.0}^{*}$                    & $-9.24_{-0.06}^{+0.05}$     \\  
3130010147 & Soft & $11.10_{-0.17}^{+0.18}$        & $<0.7$              & ${3.7}^{*}$                    & ${-0.004}^{*}$                 & $1.008_{-0.023}^{+0.024}$      & $-8.451_{-0.011}^{+0.011}$     & ${2.0}^{*}$                    & $-9.44_{-0.06}^{+0.05}$       \\
4130010101 & Hard & $10.70_{-0.20}^{+0.20}$           & $<0.27$            & ${3.0}^{*}$                    & ${-0.004}^{*}$                 & ${0.7}^{*}$                    & $<-10.0$                  & $1.69_{-0.05}^{+0.06}$         & $-8.949_{-0.007}^{+0.006}$    \\
4130010102 & Hard & $10.6_{-0.5}^{+1.2}$           & $<0.18$            & ${3.0}^{*}$                    & ${-0.004}^{*}$                 & $0.78_{-0.16}^{+0.27}$         & $<-8.9$           & $1.73_{-0.06}^{+0.06}$         & $-8.89_{-0.23}^{+0.05}$     \\  
4130010103 & Hard & $10.4_{-0.7}^{+0.7}$           & $<0.17$            & ${3.1}^{*}$                    & ${-0.004}^{*}$                 & $0.90_{-0.21}^{+0.8}$          & $<-9.1$             & $1.69_{-0.20}^{+0.09}$          & $-8.81_{-0.13}^{+0.04}$     \\  
4130010104 & Hard & $10.7_{-0.7}^{+0.9}$           & $<0.15$            & ${3.2}^{*}$                    & ${-0.004}^{*}$                 & $0.81_{-0.24}^{+0.7}$          & $<-9.0$             & $1.67_{-0.23}^{+0.05}$         & $-8.74_{-0.22}^{+0.05}$     \\  
4130010105 & Hard & $10.03_{-0.5}^{+0.19}$         & $<0.23$            & ${3.3}^{*}$                    & ${-0.004}^{*}$                 & $1.12_{-0.22}^{+0.29}$         & $-9.35_{-0.9}^{+0.27}$         & ${1.8}^{*}$                    & $-8.71_{-0.09}^{+0.06}$      \\ 
4130010106 & Hard & $10.4_{-0.3}^{+0.4}$           & $<0.17$          & ${3.3}^{*}$                    & ${-0.004}^{*}$                 & $1.08_{-0.17}^{+0.20}$          & $-8.98_{-0.12}^{+0.09}$        & ${1.8}^{*}$                    & $-8.63_{-0.06}^{+0.06}$      \\ 
4130010107 & Hard & $10.07_{-0.4}^{+0.15}$         & $<0.15$          & ${3.5}^{*}$                    & ${-0.004}^{*}$                 & $0.81_{-0.10}^{+0.13}$          & $-9.29_{-0.9}^{+0.25}$         & ${2.0}^{*}$                    & $-8.44_{-0.05}^{+0.03}$       \\
4130010108 & Hard & $10.78_{-0.29}^{+0.3}$         & $<0.20$          & ${3.5}^{*}$                    & ${-0.004}^{*}$                 & $0.89_{-0.08}^{+0.07}$         & $-8.78_{-0.07}^{+0.06}$        & ${2.0}^{*}$                    & $-8.45_{-0.04}^{+0.04}$       \\
4130010110 & Hard & $11.37_{-0.28}^{+0.3}$         & $<1.7$              & ${3.6}^{*}$                    & ${-0.004}^{*}$                 & $0.90_{-0.06}^{+0.06}$         & $-8.401_{-0.03}^{+0.028}$      & ${2.0}^{*}$                    & $-8.27_{-0.04}^{+0.04}$       \\
4130010111 & Hard & $11.26_{-0.20}^{+0.21}$         & $<0.16$            & ${3.7}^{*}$                    & ${-0.004}^{*}$                 & $1.00_{-0.05}^{+0.05}$         & $-8.353_{-0.021}^{+0.022}$     & ${2.0}^{*}$                    & $-8.190_{-0.022}^{+0.021}$    \\
4130010112 & Hard & $10.39_{-0.29}^{+0.21}$        & $<0.7$            & ${3.5}^{*}$                    & ${-0.004}^{*}$                 & $0.80_{-0.20}^{+0.06}$            & $-8.72_{-0.05}^{+0.05}$        & ${2.0}^{*}$                    & $-8.365_{-0.029}^{+0.008}$    \\
4130010113 & Hard & $10.7_{-0.4}^{+0.4}$           & $<0.7$            & ${3.5}^{*}$                    & ${-0.004}^{*}$                 & $0.93_{-0.08}^{+0.09}$         & $-8.54_{-0.05}^{+0.06}$        & ${2.0}^{*}$                    & $-8.39_{-0.06}^{+0.04}$       \\
4130010114 & Hard & $11.03_{-0.17}^{+0.17}$        & $<0.17$            & ${3.7}^{*}$                    & ${-0.004}^{*}$                 & $1.02_{-0.03}^{+0.05}$         & $-8.40_{-0.04}^{+0.07}$        & ${2.0}^{*}$                    & $-8.276_{-0.05}^{+0.028}$     \\
4130010115 & Soft & $11.60_{-0.10}^{+0.11}$         & $<0.4$            & ${4.2}^{*}$                    & ${-0.004}^{*}$                 & $1.26_{-0.06}^{+0.05}$         & $-7.980_{-0.018}^{+0.017}$     & ${2.0}^{*}$                    & $-8.023_{-0.029}^{+0.029}$    \\
4130010116 & Soft & $11.84_{-0.08}^{+0.08}$        & $<0.4$            & ${4.2}^{*}$                    & ${-0.004}^{*}$                 & $1.63_{-0.07}^{+0.07}$         & $-7.720_{-0.019}^{+0.019}$     & ${2.0}^{*}$                    & $-8.04_{-0.05}^{+0.05}$       \\
4130010118 & Soft & $11.66_{-0.12}^{+0.12}$        & $<0.5$              & ${4.2}^{*}$                    & ${-0.004}^{*}$                 & $1.59_{-0.08}^{+0.09}$         & $-7.703_{-0.021}^{+0.021}$     & ${2.0}^{*}$                    & $-8.08_{-0.07}^{+0.06}$       \\
4130010121 & Soft & $11.53_{-0.07}^{+0.07}$        & $<0.4$              & ${4.1}^{*}$                    & ${-0.004}^{*}$                 & $1.359_{-0.024}^{+0.025}$      & $-7.763_{-0.004}^{+0.004}$     & ${2.0}^{*}$                    & $-8.39_{-0.03}^{+0.03}$      \\ 
4130010122 & Soft & $11.47_{-0.11}^{+0.11}$        & $<0.7$              & ${4.1}^{*}$                    & ${-0.004}^{*}$                 & $1.34_{-0.04}^{+0.04}$         & $-7.782_{-0.006}^{+0.006}$     & ${2.0}^{*}$                    & $-8.36_{-0.04}^{+0.04}$       \\
4130010123 & Soft & $11.21_{-0.13}^{+0.14}$        & $<0.7$              & ${4.0}^{*}$                    & ${-0.004}^{*}$                 & $1.27_{-0.04}^{+0.04}$         & $-7.884_{-0.008}^{+0.008}$     & ${2.0}^{*}$                    & $-8.29_{-0.04}^{+0.04}$       \\
4130010124 & Soft & $11.28_{-0.11}^{+0.12}$        & $<0.7$            & ${4.0}^{*}$                    & ${-0.004}^{*}$                 & $1.23_{-0.03}^{+0.04}$         & $-7.885_{-0.006}^{+0.006}$     & ${2.0}^{*}$                    & $-8.254_{-0.03}^{+0.027}$     \\
4130010125 & Soft & $11.23_{-0.11}^{+0.03}$        & $<0.9$            & ${4.0}^{*}$                    & ${-0.004}^{*}$                 & $1.38_{-0.04}^{+0.04}$         & $-7.8875_{-0.0024}^{+0.003}$   & ${2.0}^{*}$                    & $-8.564_{-0.007}^{+0.006}$    \\
4130010126 & Soft & $10.90_{-0.14}^{+0.14}$        & $<0.7$            & ${4.0}^{*}$                    & ${-0.004}^{*}$                 & $1.56_{-0.09}^{+0.10}$          & $-7.923_{-0.021}^{+0.021}$     & ${2.0}^{*}$                    & $-8.36_{-0.09}^{+0.07}$       \\
4130010127 & Soft & $10.86_{-0.26}^{+0.27}$        & $<0.6$             & ${3.8}^{*}$                    & ${-0.004}^{*}$                 & $1.16_{-0.07}^{+0.06}$         & $-8.154_{-0.019}^{+0.022}$     & ${2.0}^{*}$                    & $-8.27_{-0.04}^{+0.04}$       \\
4130010128 & Soft & $10.97_{-0.20}^{+0.21}$         & $<1.1$            & ${4.0}^{*}$                    & ${-0.004}^{*}$                 & $1.46_{-0.11}^{+0.12}$         & $-8.006_{-0.025}^{+0.026}$     & ${2.0}^{*}$                    & $-8.38_{-0.10}^{+0.08}$       \\ 
6130010107 & Soft & $11.30_{-0.09}^{+0.09}$        & $<1.3$             & ${4.2}^{*}$                    & ${-0.004}^{*}$                 & $1.41_{-0.03}^{+0.04}$         & $-7.699_{-0.006}^{+0.006}$     & ${2.0}^{*}$                    & $-8.27_{-0.04}^{+0.04}$       \\
6557010102 & Soft & $11.44_{-0.04}^{+0.04}$        & $<0.4$              & ${4.2}^{*}$                    & ${-0.004}^{*}$                 & $1.478_{-0.018}^{+0.018}$      & $-7.657_{-0.003}^{+0.003}$     & ${2.0}^{*}$                    & $-8.397_{-0.029}^{+0.027}$    \\
6557010302 & Soft & $11.57_{-0.04}^{+0.04}$        & $<0.21$            & ${4.2}^{*}$                    & ${-0.004}^{*}$                 & $1.416_{-0.023}^{+0.024}$      & $-7.667_{-0.005}^{+0.005}$     & ${2.0}^{*}$                    & $-7.905_{-0.015}^{+0.013}$    \\
6130010109 & Soft & $10.57_{-0.10}^{+0.11}$         & $<0.5$              & ${4.1}^{*}$                    & ${-0.004}^{*}$                 & $1.42_{-0.07}^{+0.07}$         & $-7.893_{-0.023}^{+0.022}$     & ${2.0}^{*}$                    & $-8.06_{-0.04}^{+0.04}$       \\
5665010406 & Soft & $11.93_{-0.09}^{+0.09}$        & $<2.1$           & ${4.3}^{*}$                    & ${-0.004}^{*}$                 & $1.63_{-0.04}^{+0.04}$         & $-7.595_{-0.005}^{+0.005}$     & ${2.0}^{*}$                    & $-9.18_{-0.6}^{+0.24}$        \\
5665010407 & Soft & $11.74_{-0.04}^{+0.04}$        & $2.9_{-1.2}^{+4}$              & $4.33_{-0.18}^{+0.7}$          & $>-0.010$         & $1.562_{-0.016}^{+0.016}$      & $-7.6652_{-0.0022}^{+0.0021}$  & ${2.0}^{*}$                    & $-9.26_{-0.17}^{+0.12}$       \\
5665010408 & Soft & $11.72_{-0.09}^{+0.09}$        & $13_{-9}^{+4}$            & $>4.2$              & $-0.005_{-0.004}^{+0.004}$     & $1.50_{-0.03}^{+0.03}$         & $-7.713_{-0.004}^{+0.004}$     & ${2.0}^{*}$                    & $-9.11_{-0.25}^{+0.15}$       \\
6130010110 & Soft & $11.67_{-0.09}^{+0.09}$        & $<1.2$              & ${4.2}^{*}$                    & ${-0.004}^{*}$                 & $1.53_{-0.03}^{+0.04}$         & $-7.694_{-0.005}^{+0.005}$     & ${2.0}^{*}$                    & $-9.04_{-0.23}^{+0.15}$       \\
5665010409 & Soft & $11.64_{-0.07}^{+0.07}$        & $7_{-4}^{+9}$                & $>4.1$              & $-0.006_{-0.005}^{+0.005}$     & $1.506_{-0.025}^{+0.025}$      & $-7.742_{-0.003}^{+0.003}$     & ${2.0}^{*}$                    & $-9.48_{-0.5}^{+0.23}$       \\ 
6130010111 & Soft & $11.70_{-0.05}^{+0.05}$        & $18.3_{-8}^{+2.3}$          & $>4.6$              & $>-0.003$            & $1.466_{-0.010}^{+0.007}$       & $-7.7954_{-0.0025}^{+0.0025}$  & ${2.0}^{*}$                    & $<-10.1$           \\  
6130010112 & Soft & $11.70_{-0.05}^{+0.05}$        & $17.2_{-11}^{+2.1}$          & $>4.3$              & $>-0.005$       & $1.460_{-0.004}^{+0.007}$      & $-7.8001_{-0.0027}^{+0.0025}$  & ${2.0}^{*}$                    & $<-10.3$          \\   
6130010113 & Soft & $11.80_{-0.06}^{+0.06}$        & $14_{-7}^{+6}$            & $>4.3$              & $-0.0036_{-0.0024}^{+0.0024}$  & $1.423_{-0.017}^{+0.017}$      & $-7.8326_{-0.0025}^{+0.0026}$  & ${2.0}^{*}$                    & $-9.49_{-0.24}^{+0.15}$       \\
5665010410 & Soft & $11.86_{-0.06}^{+0.06}$        & $<2.0$              & ${4.1}^{*}$                    & ${-0.004}^{*}$                 & $1.479_{-0.020}^{+0.021}$       & $-7.7506_{-0.0028}^{+0.0028}$  & ${2.0}^{*}$                    & $-9.16_{-0.14}^{+0.11}$       \\
5665010411 & Soft & $11.88_{-0.05}^{+0.05}$        & $19.2_{-9}^{+1.9}$          & $>4.5$              & $-0.0035_{-0.0026}^{+0.0025}$  & $1.417_{-0.004}^{+0.007}$      & $-7.8489_{-0.0027}^{+0.0027}$  & ${2.0}^{*}$                    & $<-10.1$             \\
6130010117 & Soft & $12.03_{-0.07}^{+0.07}$        & $<2.2$              & ${4.1}^{*}$                    & ${-0.004}^{*}$                 & $1.417_{-0.020}^{+0.021}$         & $-7.8193_{-0.0029}^{+0.0029}$  & ${2.0}^{*}$                    & $-9.69_{-0.6}^{+0.25}$        \\
6130010118 & Soft & $11.88_{-0.07}^{+0.04}$        & $18.9_{-12}^{+2.3}$          & $>4.6$              & $>-0.006$        & $1.395_{-0.005}^{+0.009}$      & $-7.8617_{-0.003}^{+0.0022}$   & ${2.0}^{*}$                    & $<-10.3$            \\ 
6130010119 & Soft & $11.98_{-0.05}^{+0.05}$        & $10.9_{-2.7}^{+5}$          & $4.10_{-0.19}^{+0.3}$          & $-0.0028_{-0.0015}^{+0.0016}$  & $1.385_{-0.007}^{+0.007}$      & $-7.9113_{-0.0028}^{+0.0028}$  & ${2.0}^{*}$                    & $<-10.8$             \\
6130010120 & Soft & $12.04_{-0.10}^{+0.10}$           & $7.3_{-2.7}^{+2.9}$            & $3.78_{-0.22}^{+0.21}$         & $>-0.003$           & $1.381_{-0.017}^{+0.012}$      & $-7.913_{-0.005}^{+0.005}$     & ${2.0}^{*}$                    & $<-9.8$            \\ 
6130010121 & Soft & $11.83_{-0.06}^{+0.06}$        & $15.9_{-3}^{+2.9}$          & $4.07_{-0.20}^{+0.15}$          & $-0.0024_{-0.0012}^{+0.0011}$  & $1.405_{-0.008}^{+0.008}$      & $-7.937_{-0.003}^{+0.003}$     & ${2.0}^{*}$                    & $<-10.8$           \\

\end{longtable}



\begin{figure}
       \centering
       \includegraphics[width=0.53\linewidth]{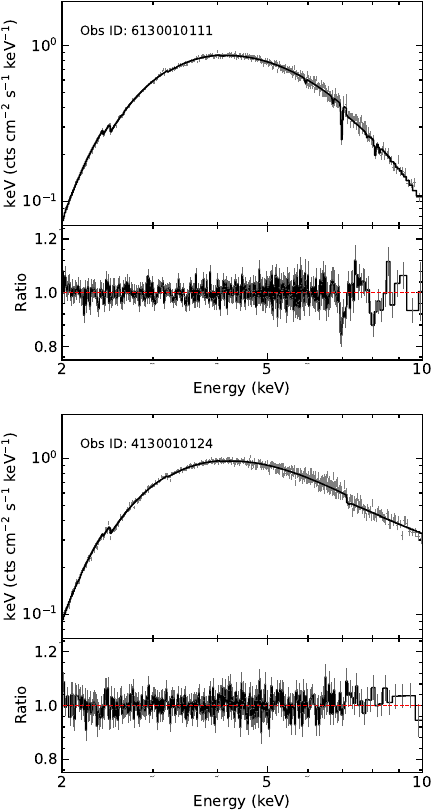} 
       \\
       \vspace{-0.3cm}
       \caption{\textbf{Supplementary Figure \thefigure~\boldmath{$|$} NICER $2-10$~keV unfolded spectrum and the fitting residuals of the two observations mentioned in Figure~1.} The fitting are conducted with the phenomenological model: tbabs $\times$ xscat $\times$ edge $\times$ (cflux $\times$ diskbb + cflux $\times$ nthcomp). The two observation have comparable flux and similar spectra, but the first observation (Obs ID: 6130010111) shows evidence of X-ray disk winds, the second observation (Obs ID: 4130010124) does not. Both spectrum and fitting residuals are shown with 1$\sigma$ errors.}
       \label{xray_spectra_fitting}
\end{figure}



\begin{figure}
       \centering
       \includegraphics[width=1.0\linewidth]{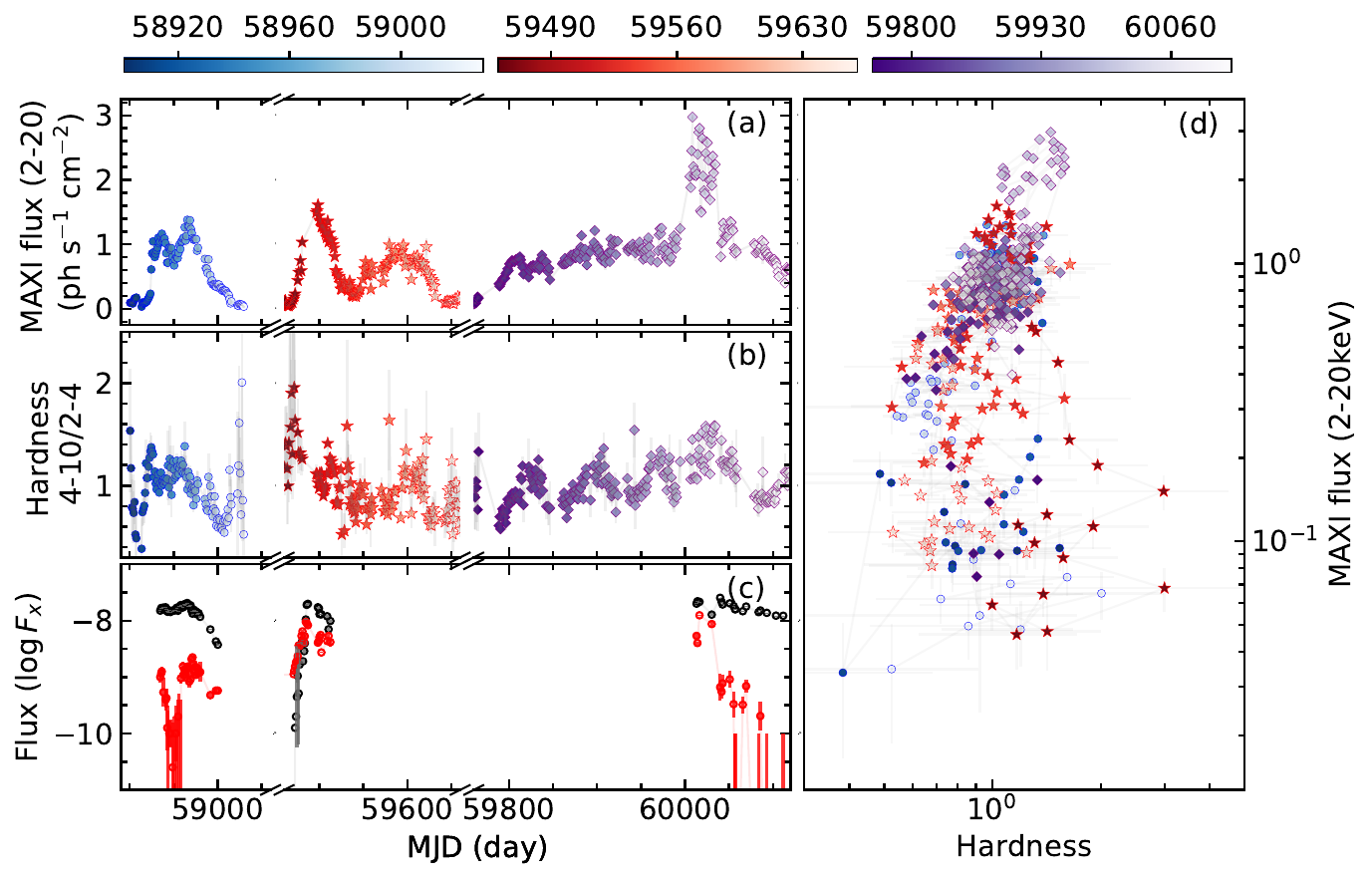} 
       \\
       \vspace{-0.3cm}
       \caption{\textbf{Supplementary Figure \thefigure~\boldmath{$|$} Monitoring of the 2020, 2021 and 2022-2023 outbursts of \src.} (a) The long term $2-20$~keV MAXI light curve. (b) The corresponding MAXI hardness, which is defined as the ratio between count rates in $4-10$~keV and $2-4$~keV bands. (c) The unabsorbed $1-10$~keV flux (in units of erg~cm$^{-2}$~s; 3$\sigma$ errors) of the disk component (dark points) and the corona component (red points) for each NICER observation, which are extracted from the best-fit phenomenological model: tbabs $\times$ xscat $\times$ edge $\times$ (cflux $\times$ diskbb + cflux $\times$ nthcomp + gauss). (d) The MAXI hardness-intensity diagram, defined as the total $2-20$~keV count rate vs. the hardness ($4-10$~keV/$2-4$~keV). For panels (a), (b) and (d), the data point colors correspond to the time of the observation to facilitate comparison, and error bars represent 1$\sigma$ confidence intervals.}
       \label{maxi_lcurve}
\end{figure}



\begin{figure}
       \centering
       \vspace{-0.3cm}
       \includegraphics[width=0.9\linewidth]{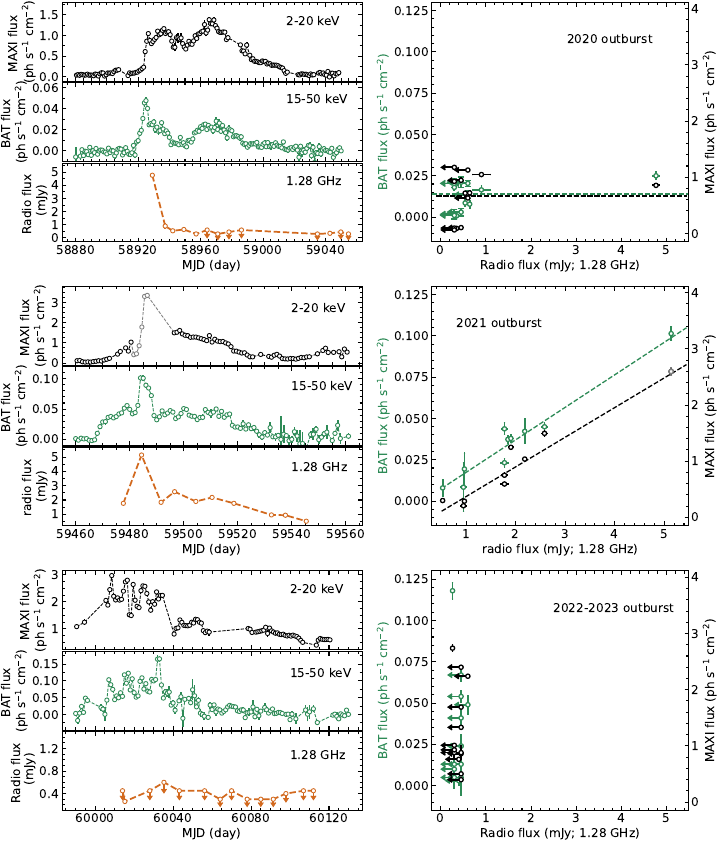}
       \vspace{-0.3cm}
       \newpage 
       \caption{\textbf{Supplementary Figure \thefigure~\boldmath{$|$} Multi-wavelength monitoring and radio--X-ray correlations for recent three outbursts.} Left panels: MAXI/GSC (2–20~keV), Swift/BAT (15–50~keV), and MeerKAT (1.28~GHz) light curves, shown from top to bottom, respectively. Right: Correlations between the radio flux density and the simultaneous X-ray fluxes from Swift/BAT (green) and MAXI/GSC (black). For 2021 outburst, the grey point represent MAXI fluxes scaled from simultaneous NICER observations. Error bars indicate 1$\sigma$ confidence intervals, and 3$\sigma$ upper limits are shown for radio flux densities in cases of non-detection.}
       \label{xray_counts_vs_radio_2}
\end{figure}



\begin{figure}
       \centering
       \includegraphics[width=0.9\linewidth]{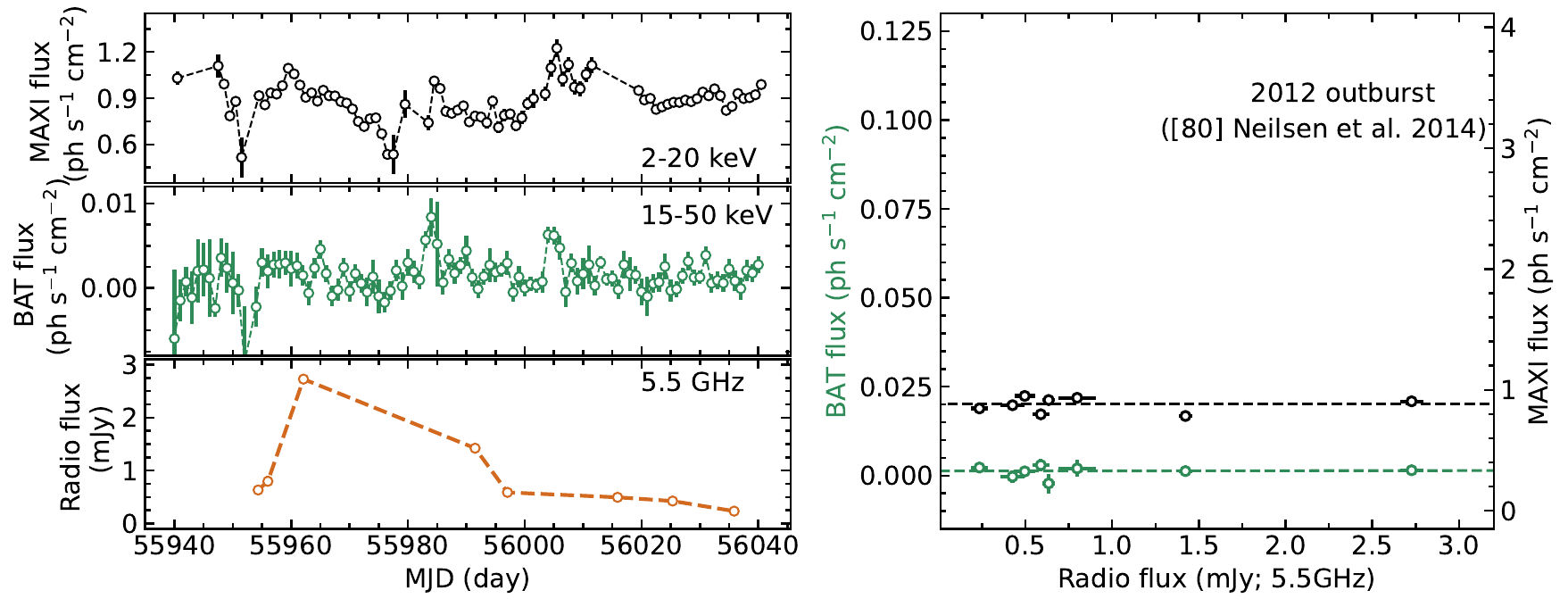}
       \\
       \vspace{-0.3cm}
       \caption{\textbf{Supplementary Figure \thefigure~\boldmath{$|$} Multi-wavelength monitoring and radio--X-ray correlations for 2012 outburst.} Left panels show MAXI/GSC (2–20~keV), Swift/BAT (15–50~keV), and ATCA (5.5~GHz) light curves, respectively. Right panel depicts correlation between the radio flux density and the simultaneous X-ray fluxes in Swift/BAT (green) and MAXI/GSC (black) energy bands. Error bars correspond to 1$\sigma$ confidence intervals.}
       \label{xray_counts_vs_radio_12}
\end{figure}



\begin{figure}
       \centering
       \includegraphics[width=0.92\linewidth]{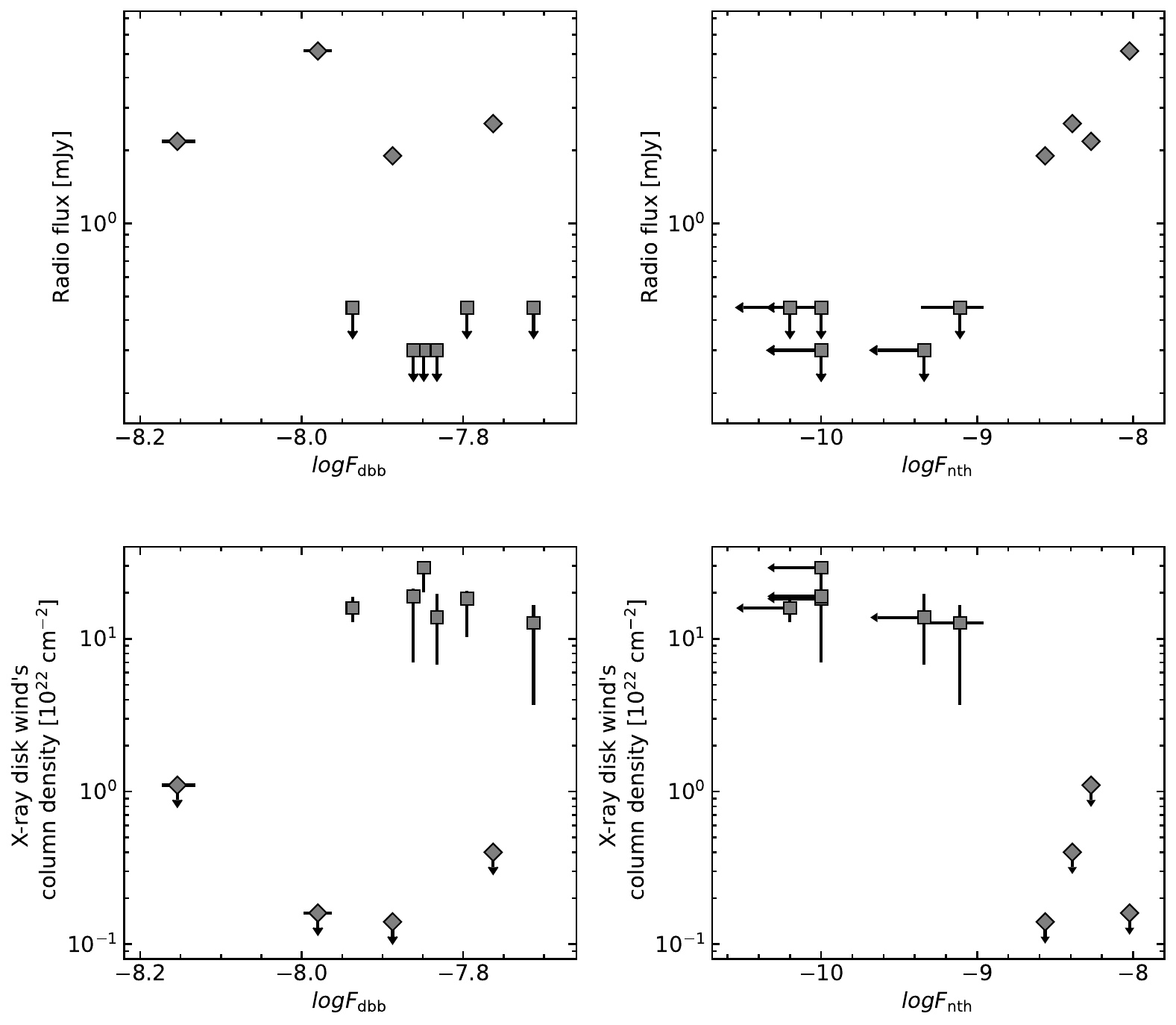} 
       \\
       \vspace{-0.3cm}
       \caption{\textbf{Supplementary Figure \thefigure~\boldmath{$|$} Summary of the relations between disk fluxes $F_{\rm dbb}$, non-thermal fluxes $F_{\rm nth}$, radio flux densities, and disk wind column densities.} Diamonds indicate observations with jets but no X-ray disk winds, while squares mark observations with X-ray disk winds but no jets. In both jet-active and wind-active states, the disk fluxes are comparable. Higher radio flux densities are generally associated with enhanced coronal emission. Error bars indicate 3$\sigma$ confidence intervals, with 3$\sigma$ upper limits shown for non-detections or unconstrained measurements.}
       \label{radio_wind_flux}
\end{figure}


\begin{figure}
       \centering
       \includegraphics[width=0.9\linewidth]{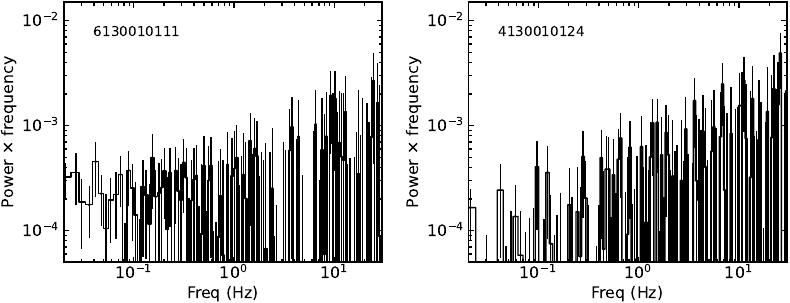} 
       \\
       \vspace{-0.3cm}
       \caption{\textbf{Supplementary Figure \thefigure~\boldmath{$|$} Power-density spectra of two typical NICER observations.} The spectra are shown with 1$\sigma$ confidence interval error bars, and scaled with frequency to illustrate the potential (Quasi-periodic oscillations) QPOs rather than the low-frequency red noise. The left and right panels represent the results of X-ray winds active observation (ID:~6130010111), and radio active observation (ID:~4130010124), respectively. X-ray winds active observation and jets active observation share consistent power density spectrum.}
       \label{pds_comparison}
\end{figure}





\begin{figure}
       \centering
       \includegraphics[width=0.6\linewidth]{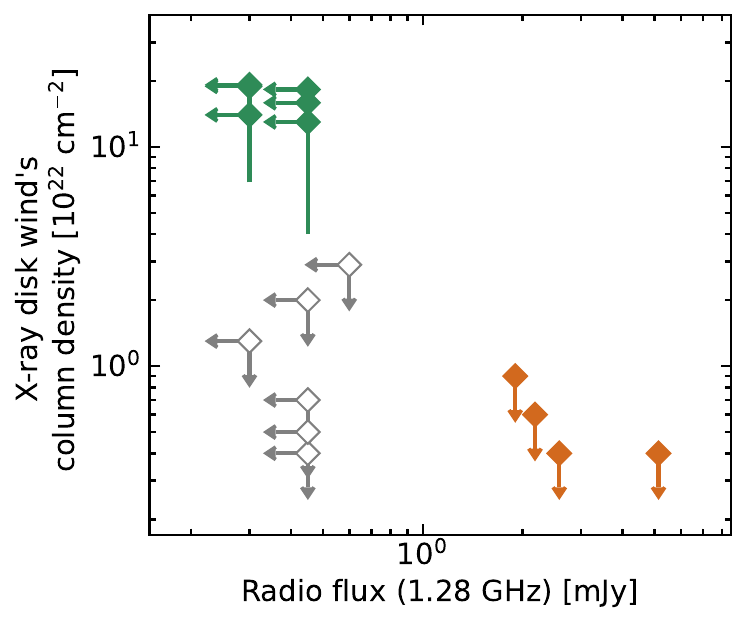} 
       \\
       \vspace{-0.3cm}
       \caption{\textbf{Supplementary Figure \thefigure~\boldmath{$|$} The X-ray wind’s column density versus radio flux density for the simultaneous observations.} The dark-yellow points represent observations where the jets were detected but no X-ray disk winds, and the green points represent observations where the X-ray disk winds were detected but no jets. The gray points represent observations where neither X-ray winds nor jets were detected. The X-ray wind’s column density are shown with 3$\sigma$ confidence interval error bars and 3$\sigma$ upper limits for non-detections; while the radio flux density are shown with 1$\sigma$ confidence interval error bars and 3$\sigma$ upper limits for non-detections.}
       \label{correlaton_wind_radio}
\end{figure}



\end{document}